\documentclass[twocolumn]{aastex631}
\usepackage{newtxtext,newtxmath}
\usepackage[T1]{fontenc}
\usepackage{graphicx}	
\usepackage{amsmath}	
\usepackage{amssymb}	
\usepackage{tabularx}

\usepackage{xcolor}
\usepackage{hyperref}
\usepackage{subfigure}
\usepackage{soul}

\shorttitle{MaDCoWS2: Evolution of SFR and Stellar Masses in Groups \& Clusters since $z \sim 2.5$}
\shortauthors{Trudeau et al.}

\graphicspath{{./}{figures/}}

\turnoffedit
\begin{document}


\title{The Massive and Distant Clusters of WISE Survey 2: A Stacking Analysis Investigating the Evolution of Star Formation Rates and Stellar Masses in Groups and Clusters}


\author[0000-0003-3428-1106]{A. Trudeau}
\email{arianetrudeau@ufl.edu}
\affiliation{Department of Astronomy, University of Florida, 211 Bryant Space Center, Gainesville, FL 32611, USA}

\author[0000-0002-0933-8601]{Anthony H. Gonzalez}
\affiliation{Department of Astronomy, University of Florida, 211 Bryant Space Center, Gainesville, FL 32611, USA}

\author[0000-0001-7027-2202]{K. Thongkham}
\affiliation{Department of Astronomy, University of Florida, 211 Bryant Space Center, Gainesville, FL 32611, USA}

\author[0000-0003-3004-9596]{Kyoung-Soo Lee}
\affiliation{Department of Physics and Astronomy, Purdue University, 525 Northwestern Avenue, West Lafayette, IN 47907, USA}

\author[0000-0002-8909-8782]{Stacey Alberts}
\affiliation{Steward Observatory, University of Arizona, 933 N. Cherry Avenue, Tucson, AZ 85721, USA}

\author[0000-0002-4208-798X]{M. Brodwin}
\affiliation{Department of Physics and Astronomy, University of Missouri, 5110 Rockhill Road, Kansas City, MO 64110, USA}

\author[0000-0002-7898-7664]{Thomas Connor}
\affiliation{Center for Astrophysics $\vert$\ Harvard\ \&\ Smithsonian, 60 Garden St., Cambridge, MA 02138, USA}

\author{Peter R. M. Eisenhardt}
\affiliation{Jet Propulsion Laboratory, California Institute of Technology, 4800 Oak Grove Dr., Pasadena, CA 91109, USA}

\author[0000-0001-9793-5416]{Emily Moravec}
\affiliation{Green Bank Observatory, P.O. Box 2, Green Bank, WV 24944}

\author[0009-0006-2340-1845]{Eshwar Puvvada}
\affiliation{Department of Physics and Astronomy, Purdue University, 525 Northwestern Avenue, West Lafayette, IN 47907, USA}

\author[0000-0003-0122-0841]{S. A. Stanford}
\affiliation{Department of Physics, University of California, One Shields Avenue, Davis, CA, 95616, USA}


\begin{abstract}
The evolution of galaxies depends on their masses and local environments; understanding when and how environmental quenching starts to operate remains a challenge. Furthermore, studies of the high-redshift regime have been limited to massive cluster members, owing to sensitivity limits or small fields of views when the sensitivity is sufficient, intrinsically biasing the picture of cluster evolution. In this work, we use stacking to investigate the average star formation history of more than 10,000 groups and clusters drawn from the Massive and Distant Clusters of WISE Survey 2 (MaDCoWS2). Our analysis covers near ultraviolet to far infrared wavelengths, for galaxy overdensities at $0.5 \lesssim z \lesssim 2.54$. We employ SED fitting to measure the specific star formation rates (sSFR) in four annular apertures with radii between 0 and 1000 kpc. At $z \gtrsim 1.6$, the average sSFR evolves similarly to the field in both the core and the cluster outskirts. Between $\overline{z} = 1.60$ and $\overline{z} = 1.35$, the sSFR in the core drops sharply, and continues to fall relative to the field sSFR at lower redshifts. We interpret this change as evidence that the impact of environmental quenching dramatically increases at $z \sim 1.5$, with the short time span of the transition suggesting that the environmental quenching mechanism dominant at this redshift operates on a rapid timescale. We find indications that the sSFR may decrease with increasing host halo mass, but lower-scatter mass tracers than the signal-to-noise ratio (S\slash N) are needed to confirm this relationship.
\end{abstract} 


\section{Introduction} \label{sec_intro}

Galaxy properties correlate with their stellar masses and the environment in which they evolve. Irrespective of environment, star formation in massive galaxies tends to cease earlier than in less massive ones \citep[e.g.][]{kauffmann_dependence_2003,peng_mass_2010,nelson_illustris_2015}, while denser environments also tend to host a larger fraction of quiescent galaxies than in the field \citep[e.g.][]{dressler_galaxy_1980,balogh_galaxy_2004,peng_mass_2010,peng_mass_2012,kawinwanichakij_effect_2017,pintos-castro_evolution_2019,van_der_burg_gogreen_2020}. Processes that tend to halt star formation in massive galaxies are referred to as ``mass'' or ``internal'' quenching while those pertaining to the local density are referred to as ``environmental quenching'' \citep[e.g.][]{peng_mass_2010}.

The most commonly cited environmental processes in the literature are ram pressure stripping, starvation, overconsumption, and harassment. Ram pressure stripping refers to infalling galaxies being stripped of their gas by the pressure of the intergalactic medium of a massive halo \citep[e.g.][]{gunn_infall_1972,poggianti_star_1999}, while starvation refers to processes which either partially remove the galaxy diffuse halo of hot gas \citep{larson_evolution_1980,balogh_origin_2000,bekki_passive_2002} or prevent inflows \citep{balogh_origin_2000,kawata_strangulation_2008,baxter_when_2023}. Without replenishment, the galaxy's gas reservoir is then exhausted by star formation. Starvation usually occurs on a timescale of several gigayears, but outflows generated by AGN feedback or vigorous star formation can accelerate the exhaustion the gas reservoir to a timescale of $\sim$1 Gyr \citep{brodwin_era_2013,mcgee_overconsumption_2014,alberts_star_2016,balogh_evidence_2016,van_der_burg_gogreen_2020}. This process is sometimes named "overconsumption" \citep[e.g.][]{mcgee_overconsumption_2014,balogh_evidence_2016} to differentiate it from the classical definition of starvation which occurs on a much longer timescale. Finally, harassment is the removal of gas and the change of morphology driven by multiple close, high-speed encounters with other galaxies, combined with the effect of the cluster potential well \citep{farouki_computer_1981,moore_galaxy_1996,moore_survival_1999,boselli_environmental_2006,bialas_occurrence_2015}. Harassment efficiency depends on many parameters, including the orbit, initial shape, surface brightness, and orientation of the galaxy.

The importance of environmental quenching in the evolution of $z \lesssim 1$ groups and clusters is well established \citep[e.g.][]{dressler_galaxy_1980,peng_mass_2010,pintos-castro_evolution_2019}, with recent works trying to establish the dominant mechanism \citep[e.g.][]{boselli_quenching_2016,boselli_virgo_2020,boselli_virgo_2023,rodriguez-munoz_quantifying_2019,kim_gradual_2023}. The situation is different at high-redshift, where the importance of environmental quenching remains unclear. There is an ongoing debate about the onset of environmental quenching in clusters, with some studies suggesting that environmental quenching has little effect before $z \sim 1.5$ \citep[e.g.][]{brodwin_era_2013,alberts_star_2016,nantais_evidence_2017,nantais_halpha_2020} while others \citep[e.g.][]{cooke_submillimetre_2019,lemaux_persistence_2019,strazzullo_galaxy_2019,van_der_burg_gogreen_2020} conclude that environmental quenching is important at $z > 1.5$. Both perspectives are supported by observations of individual $z \gtrsim 1.5$ clusters and protoclusters exhibiting a wide variety of states. Some high-redshift clusters appear to be mostly quiescent \citep[e.g.][]{gobat_mature_2011,andreon_jkcs_2014,willis_spectroscopic_2020} while others are still highly star-forming in their cores \citep[e.g.][]{miley_spiderweb_2006,webb_extreme_2015,damato_discovery_2020}.

An important factor in this debate is the challenge of detecting high-redshift clusters. There are only a handful of clusters at $z \gtrsim 1.7$ per survey \citep[e.g. the Spitzer Adaptation of the Red-Sequence Cluster Survey, \citealt{wilson_clusters_2006,wilson_spectroscopic_2009}, the South Pole Telescope Sunyaev-Zel'dovich Survey, \citealt{bleem_galaxy_2015}, and the XXL survey,][]{pierre_xxl_2016}, which means that most studies attempting to characterize high-redshift clusters are either subject to small sample statistics and/or heterogeneous samples.

The depth of existing imaging is an additional challenge. Most high-redshift studies probe cluster members at stellar masses $\gtrsim \mathrm{10^{10}~M_\odot}$ \citep[e.g.][]{brodwin_era_2013,alberts_evolution_2014,alberts_star_2016,nantais_stellar_2016,nantais_evidence_2017,wagner_evolution_2017,strazzullo_galaxy_2019,van_der_burg_gogreen_2020}. However, \citet{alberts_measuring_2021} found that galaxies with stellar masses higher than 1.26$\times \mathrm{10^{10}~M_\odot}$ contribute only 20 to 30\% of the total far infrared luminosity of a cluster. Given that quenching is mass-dependent, this raises the question of whether the behavior of massive galaxies is representative of the cluster as a whole \citep[see also][]{popescu_tracing_2023}.

Stacking on cluster positions is a way of mitigating these issues. ``Stacking'' is a statistical method that can be defined as computing the sum or the average light of a set of astronomical objects \citep[e.g.][]{kelly_60_1990,dole_cosmic_2006}. The increase of signal-to-noise ratio generated by stacking corresponds approximately to the square root of the number of stacked images \citep{kelly_60_1990,garn_radio_2009,bourne_evolution_2011}. Provided that a sufficient number of images are included, stacking can thus be used to study the statistical contribution of cluster components difficult to detect, such as low-mass galaxies or intracluster light \citep[e.g.][]{montier_dust_2005,zibetti_intergalactic_2005,gutierrez_dust_2017,alberts_measuring_2021,mckinney_measuring_2022,popescu_tracing_2023}.

In this work, we present a multiwavelength stacking analysis of the spectral energy distribution (SED) and corresponding star formation history of more than 10,000 massive groups and clusters at $0.5 \leq z \leq 2.54$. This article is structured as follows: Section \ref{sec_data} presents the cluster sample and datasets used; the stacking methodology is explained in Section \ref{sec_processing}. Section \ref{sec_results} describes the initial results and SED fitting while Section \ref{sec_discussion} discusses the star formation histories of the stacks and their implications. Our main conclusions are summarised in Section \ref{sec_conclusion}. Magnitudes are in the AB system and we assume a \citet{chabrier_galactic_2003} initial mass function. We adopt the \citet{collaboration_planck_2020} $\Lambda$CDM cosmology as implemented in {\tt astropy.cosmology} with $\Omega_m =0.31$ and $H_0 = 67.7$ km s$^{-1}$ Mpc$^{-1}$. All size scales are physical unless noted otherwise.

\section{Data}\label{sec_data}

\begin{figure}

    \centering
    
    \includegraphics[width=0.9\columnwidth]{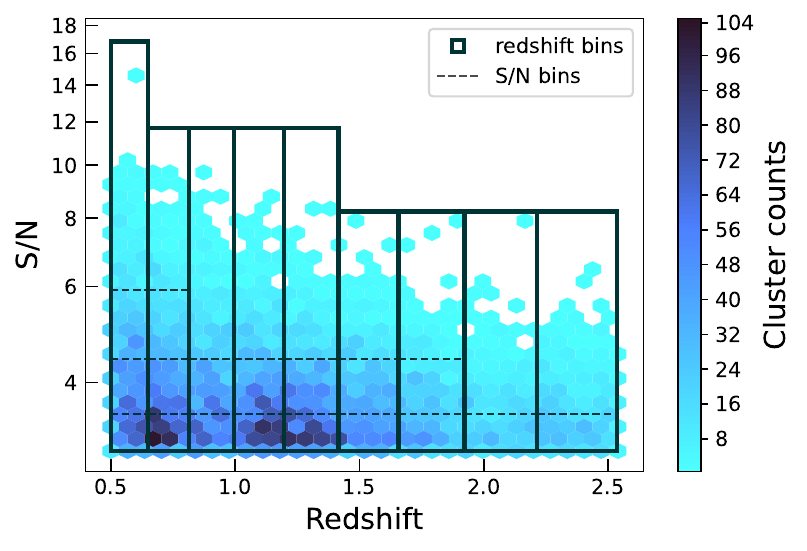}
  
    \includegraphics[width=0.9\columnwidth]{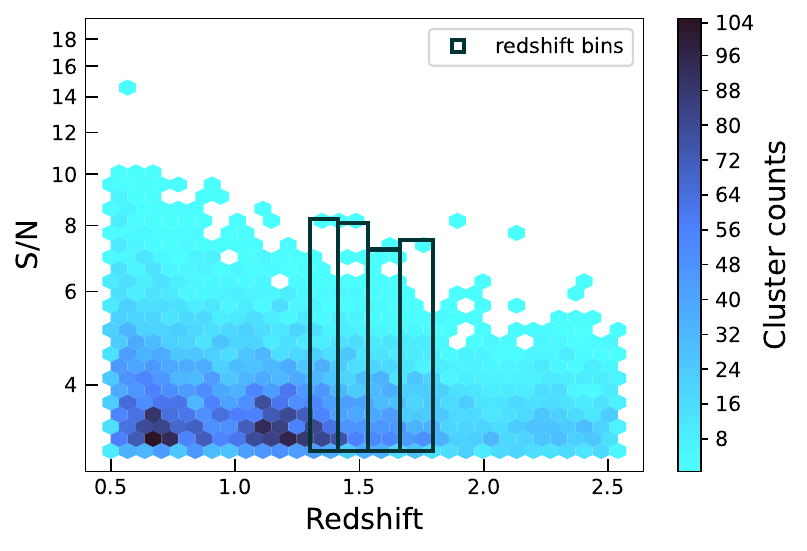}
    
    \caption{The photometric redshift and S\slash N subdivisions used in this work. The cluster sample is represented by hexagons, color-coded by counts. Top: The main redshift divisions used in this work, indicated as rectangles delimited by full lines, with the height of the rectangle giving the maximum S/N in a redshift bin. The dashed lines show the optional S\slash N subdivisions. Bottom: The tighter redshift binning used to study the sample evolution between $z \sim 1.3$ and $z \sim 1.8$.}

    \label{fig_binning}
    
\end{figure}


\begin{deluxetable*}{cccccc}
\tablecaption{Number of clusters in each redshift bin and S\slash N subdivisions.``Merged'' means that a S\slash N subdivision did not contain 75 clusters and had to be merged with a lower S\slash N subdivision. ``Total'' refer to the number of clusters in a redshift bin when the S\slash N subdivisions are turned off.}\label{tab_binning}
\tablehead{
\colhead{Mean redshift} & & \colhead{Cluster count} & & & \colhead{Total count}\\
& \colhead{3.0 < S\slash N < 3.5} & \colhead{3.5 < S\slash N < 4.4} & \colhead{4.4 < S\slash N < 5.9} & \colhead{S\slash N > 5.9} & 
}
\startdata
$\overline{0.58}$ & 403 & 462 & 324 & 143 & 1332\\
$\overline{0.72}$ & 584 & 573 & 287 & 93 & 1537\\
$\overline{0.90}$ & 470 & 480 & 292 & merged & 1242\\
$\overline{1.10}$ & 679 & 607 & 292 & merged & 1578\\
$\overline{1.30}$ & 707 & 626 & 220 & merged & 1553\\
$\overline{1.53}$ & 530 & 466 & 180 & merged & 1176\\
$\overline{1.77}$ & 380 & 303 & 86 & merged & 769\\
$\overline{2.07}$ & 298 & 249 & merged & merged & 547\\
$\overline{2.365}$ & 329 & 290 & merged & merged & 619\\
\enddata
\end{deluxetable*}


\begin{deluxetable}{cc}
\tablecaption{Number of clusters in each redshift bin for the subsample between $z \sim 1.3$ and $z \sim 1.8$}\label{tab_binning_subset}
\tablehead{
\colhead{Mean redshift} & \colhead{Cluster counts}
}
\startdata
$\overline{1.35}$ & 784\\
$\overline{1.47}$ & 607\\
$\overline{1.60}$ & 609\\
$\overline{1.73}$ & 450\\
\enddata
\end{deluxetable}

In this work, we use data from the \textit{Wide-field Infrared Survey Explorer (WISE)}, \textit{Herschel Space Observatory} and the \textit{Galaxy Evolution Explorer (GALEX)} to study the star formation history of clusters of galaxies. Our analysis focuses on the overlap between the first data release of the Massive and Distant Clusters of WISE Survey 2 \citep[MaDCoWS2; see][]{thongkham_massive_2024} and the Herschel Astrophysical Terahertz Large Area Survey \citep[H-ATLAS; see][]{valiante_herschel-atlas_2016,bourne_herschel-atlas_2016,smith_herschel-atlas_2017,maddox_herschel-atlas_2018}, which corresponds roughly to the three equatorial Galaxy and Mass Assembly (GAMA) fields \citep{driver_galaxy_2011}.

\subsection{Cluster catalog \& sample}\label{ssec_madcows2}

MaDCoWS2 is an optical and near-infrared survey designed to generate a galaxy cluster sample up to a redshift of $z \sim 2$, within the Dark Energy Camera (DECam) Legacy Survey footprint \citep[DECaLS; see][]{flaugher_dark_2015,dey_overview_2019}. A detailed explanation of the cluster detection and catalog construction is given in \citet{thongkham_massive_2024} and we provide only a brief overview here. The first step in building the catalog is to compute photometric redshift probability distribution functions \citep[PDF; see][]{brodwin_photometric_2006} using the W1 and W2 data drawn from the CatWISE2020 catalog and the $grz$ bands of the DECaLS catalog \citep{dey_overview_2019,eisenhardt_catwise_2020,marocco_catwise2020_2021}. Results are then processed by the PZWav algorithm \citep[e.g.][]{euclid_collaboration_euclid_2019,werner_s-plus_2023}, which generates a density map for each redshift between 0.1 and 3.0, in steps of $\Delta z = 0.06$. Each pixel in the density map has a size of 12 arcsec. The maps are then convolved with a difference-of-Gaussians kernel with inner and outer standard deviations of 400 and 2000 kpc \citep{thongkham_massive_2024}. The detected overdensities are classified by their redshift and signal-to-noise ratios (S\slash N). The S\slash N correlates with the mass and redshift \citep{thongkham_massive_2024} and is thus used as a cluster mass proxy in our analysis. Although the publicly available MaDCoWS2 catalog only contains $z \lesssim 2$ overdensities with an S\slash N of 5 or more, this work uses a catalog with less conservative cuts. We consider every overdensity with an S\slash N of more than 3 as a ``cluster.''
While the \citet{thongkham_massive_2024} mass calibrations do not extend to the low $S\slash N$ or highest redshift regimes probed by this study, we expect that at S\slash N$\sim 3$ we are probing low mass groups at $z \sim 0.5$ and poor clusters at $z \sim 2$. We also expand the photometric redshift range to cover the $0.5 \leq z \leq 2.54$ redshift interval, for a total sample of 10,353 MaDCoWS2 candidate clusters with H-ATLAS coverage (compared to 1024 in the public catalogue). To accommodate our recursive redshift binning (see below), we choose to cut our sample at $z \sim 2.54$ rather than at $z \sim 2.5$.

We define the redshift binning recursively. Starting at a photometric redshift of 0.5, our bin size is determined by $\Delta z = 0.1 (1 + z_{min})$ where $z_{min}$ is the lower limit of a given photometric redshift bin. The $1 + z_{min}$ factor accounts for the increased uncertainty of the photometric redshifts at higher redshift. In addition, given the unknown purity of the sample beyond $z \sim 2$, we denote the highest redshift bin ($2.21 \leq z < 2.54$) by open symbols in later Figures.

Although most of our analysis focuses on redshift evolution, we introduce S\slash N subdivisions as a halo mass proxy to investigate its impact on the star formation history. These subdivisions are also computed recursively, with larger intervals for higher S\slash N, to take into account the clusters mass function \citep[e.g.][]{white_core_1978,jenkins_mass_2001,tinker_toward_2008}. Furthermore, we require each S\slash N bin to contain at least 75 clusters; S\slash N bins that do not fulfill this requirement are merged with lower S\slash N bins until they do. The top panel of Figure \ref{fig_binning} presents the final redshift binning and the S\slash N subdivisions. The bottom panel of Figure \ref{fig_binning} shows an alternative binning used to assess the evolution of the cluster sSFR with the projected radii over the $1.3 < z \lesssim 1.8$ redshift range. In this scheme, the bins begin at $z_{min} = 1.3$ and are sized by $\Delta z = 0.05 (1 + z_{min})$. The cluster counts in each bin for both binning schemes are given in Tables \ref{tab_binning} and \ref{tab_binning_subset}.

\subsection{WISE images}\label{ssec_wise_data} WISE \citep{wright_wide-field_2010} is a space-based observatory that conducted an all-sky survey in 2010 in four infrared bands centered at 3.4 $\mu$m, 4.5 $\mu$m, 12 $\mu$m and 22 $\mu$m (usually designated as W1 to W4). After 9 month of operations, WISE's reserve of cryogenic coolant was depleted. WISE continued to survey the sky until the spacecraft was placed in a hibernation state in February 2011, having completed a second coverage of the sky in W1 and W2. The satellite was reactivated in 2013 and renamed the Near-Earth Object Wide-field Infrared Survey Explorer \citep[NEOWISE;][]{mainzer_initial_2014}. In December 2013, NEOWISE resumed surveying the sky in W1 and W2 every 6 months, and survey operations are expected to conclude due to orbit decay in July 2024, after more than 23 sky coverages in W1 and W2 (J. Hunt private communication, 2024). The MaDCoWS2 cluster catalog uses CatWISE2020 photometry \citep{marocco_catwise2020_2021} measured from coadded WISE and NEOWISE images created by the unWISE team \citep{lang_unwise_2014,meisner_full-depth_2017} using 12 sky coverages.

Rather than using the images made by the WISE team \citep{masci_awaic_2009}, our analysis relies on the unblurred WISE coadded images (unWISE), made by \citet{lang_unwise_2014}. Unlike the WISE team images, the unWISE coadds have not been convolved with the PSF, thus preserving their full resolution. For W1 and W2, we use the coadded images generated with all the WISE observations taken prior to or during the seventh year of the NEOWISE mission, while the W3 coadds are based on the observations obtained before the cryogen was exhausted. We do not use the W4 coadds, as at the redshifts considered in this study, they probe the polycyclic aromatic hydrocarbon features situated between 6.2 and 12.7 microns inclusively \citep[PAH;][]{farrah_nature_2008}. The complexity of these features, together with the uncertainty in redshift of individual clusters make the result of stacking difficult to predict and to model for most SED fitting codes \citep{alberts_measuring_2021}. For the same reason, we restrict our use of the W3 stacks to $z > 1.66$. Below this redshift, W3 images and profiles are shown for illustrative purposes only. We use the publicly available unWISE cutout service\footnote{See \url{http://unwise.me/imgsearch/}} to download science and standard deviation cutouts, and the Legacy Survey viewer cutout service to get cutouts of the masks generated for the unWISE coadds \footnote{\url{http://legacysurvey.org/viewer}} (see Section \ref{sssec_wise_pipeline} for more details on these masks).

\subsection{Herschel H-ATLAS maps}\label{ssec_herschel_data}

The H-ATLAS survey is a 660 deg$^2$ far-infrared survey conducted by the Herschel Space Observatory \citep{eales_herschel_2010,smith_herschel-atlas_2017}. It consists of five fields observed with the Photodetector Array Camera \citep[PACS:][]{poglitsch_photodetector_2010} and Spectrometer and the Spectral and Photometric Imaging Receiver cameras \citep[SPIRE:][]{griffin_herschel-spire_2010}, for a total of five bands, centered on 100, 160, 250, 350, and 500 $\mu$m. The GAMA fields, totaling 161.6 deg$^2$ \citep{valiante_herschel-atlas_2016}, lie within the first MaDCoWS2 data release. Each position in the GAMA fields is covered by at least two scans in perpendicular directions \citep[see][]{valiante_herschel-atlas_2016}. As scans overlap, some positions are covered by four scans.

H-ATLAS SPIRE data have been processed with the Herschel Interactive Pipeline Environment version 8 \citep[HIPE:][]{ott_herschel_2010}. However, the PACS data reduction was based on the work of \citet{ibar_h-atlas_2010}, which departs from the standard HIPE pipeline in several significant ways \citep{valiante_herschel-atlas_2016}.

Publicly available H-ATLAS SPIRE products for the three GAMA fields\footnote{See \url{https://www.h-atlas.org/public-data/download}} consist of the level 3 ``raw'' maps (i.e. no further processing beyond the standard HIPE pipeline and the map-making algorithm); the background-subtracted maps and the background-subtracted, smoothed maps. In the background-subtracted map, dust emission more extended than 1.5 arcmin was subtracted using the {\tt Nebuliser} algorithm; the smoothed maps were convolved with a matched filter \citep{chapin_joint_2011} to further filter out extended emission. Masks, instrumental noise, and filtered noise maps are also available. The SPIRE data used in this work consist of the raw maps, the instrumental noise maps, and the masks. The angular scale used to model the background of the SPIRE maps is likely to remove any trace of extended, diffuse components \citep[such as an eventual contribution from dust in the intracluster medium; see e.g.][]{dwek_infrared_1990,bianchi_herschel_2017,gutierrez_dust_2017,alberts_measuring_2021}. For this reason, we choose to use the raw maps rather than the background-subtracted ones.

The only available PACS maps are the background-subtracted maps, in which any emission more extended than 4 arcmin was removed, and the maps showing the number of scans used to create the PACS science maps. All of the available PACS products are used.

\subsection{GALEX Medium Imaging Survey}\label{ssec_GALEX_data}
GALEX, launched in 2003 and decommissioned in 2013 \citep{martin_galaxy_2005,bianchi_revised_2017}, observed most of the sky in the far ultraviolet (FUV; 1344 to 1786 \AA) and near ultraviolet \citep[NUV; 1771 to 2831 \AA, see][]{morrissey_calibration_2007} bands. In this work, we use the NUV data from the Medium Imaging survey (MIS), as the FUV band probes the Lyman break or blueward for all redshifts considered in this work. We restrict ourselves to MIS observations because they correspond to the deepest dataset available for the GAMA field. The All-Sky Imaging observations offer only marginally better coverage, and for typical exposure times (1500 seconds for MIS and 100 seconds for AIS), \citet{morrissey_calibration_2007} MIS is $\sim$1.9 magnitudes deeper.

Given the gaps in the coverage, we did not apply any further cuts in the sample: we downloaded every MIS observation covering the GAMA fields from the STScI archives, using the {\tt astroquery} package \citep{ginsburg_astroquery_2019}. The downloaded pointings have a median exposure time of 1768 seconds with a standard deviation of 705 seconds. The data products used in our analysis are the intensity maps, the high-resolution relative response maps, and the segmentation maps generated by {\tt SExtractor} \citep{bertin_sextractor_1996}\footnote{See \url{http://www.galex.caltech.edu/wiki/Public:Documentation} chapter 5 for a more comprehensive description of the available data products}. All the GALEX data used in this paper can be found in MAST: \dataset[10.17909/0ghf-pw79]{http://dx.doi.org/10.17909/0ghf-pw79}.

\section{Data processing and stacks}\label{sec_processing}

We generate mean stacks weighted by the inverse of the variance \citep{alberts_evolution_2014}, following a method loosely based on \citet{alberts_measuring_2021}. The dimensions of our stacks are approximately 15.9 arcmin $\times$ 15.9 arcmin, with an odd number of pixels per side, to facilitate the comparison with a Navarro-Frenk-White profile \citep[hereafter NFW;][]{navarro_structure_1996} in Section \ref{ssec_aperture_correction}. At $z = 0.5$, the lowest redshift considered, 15.9 arcmin corresponds to 6 Mpc.

\subsection{WISE and GALEX stacking method}\label{ssec_wise_galex_stacking}
Image processing for WISE and GALEX is similar because both data sets are available in tiles (or pointings for GALEX) and have point spread functions (PSFs) with FWHM between 5 to 6.5 arcmin. This is small enough to resolve nearby galaxies but large enough that most $z \geq 0.5$ galaxies are point sources. The stacking method for these datasets is as follows:
\begin{enumerate}
    \item Retrieval of the science image tile(s). Cutouts for more than one tile are occasionally necessary to reconstruct the cluster image. We will refer to these as ``partial cutouts'' in the following subsections.
    \item Computation or retrieval of the sigma maps (i.e. the error maps). Like the science images, more than one sigma map might be necessary to reconstruct the variance of the cluster image.
    \item Retrieval of the initial masks of bright objects.
    \item Removal of eventual background gradients in each (partial) cutout. This is done by masking all detections and fitting a plane ($f_{bck} = Ax + By + C$ where x and y are the pixel positions) to the remaining pixels. The plane is then subtracted from the science image.
    \item Masking of stars and of bright foreground galaxies on the science cutouts. A bright galaxy is defined as a galaxy with a flux greater than ten times the characteristic luminosity at the cluster redshift (i.e. > 10 L$_*$). This liberal limit was chosen to avoid accidentally masking prominent brightest cluster galaxies. To ensure that masked objects are not used in the weighted mean, we replace their pixel values by NaN.
    \item Merging of partial cutouts wherever necessary. Science images are co-added; sigma maps are added in quadrature.
    \item Masking of photometric non-members. A non-member is defined as a galaxy for which the redshift interval created by the mode of the PDF, plus or minus three times the PDF standard deviation does not include the cluster photometric redshift. Three times the PDF standard deviation was empirically determined to be the best value to preserve the stack flux while minimizing its uncertainty.
    \item Sigma maps are inverted and squared to generate weight maps (1/$\sigma^2$).
    \item Science images and their associated weight maps are rotated by a random integer multiple of 90\textdegree~to circularize elliptical PSFs before being stacked.
\end{enumerate}

\subsubsection{WISE image processing}\label{sssec_wise_pipeline}

The unWISE cutout service generates as many science and sigma images as there are tiles. If the requested cutout overlaps with two tiles, then two science cutouts will be generated, corresponding to the coverage provided by each tile. We do not keep every partial cutout: we first check the dimensions of each partial cutout and omit any with a dimension smaller than the overlap between two tiles which is about 64 pixels (176 arcsec). Remaining cutouts are then organised by wavelengths and their fluxes converted into Janskys.

Every WISE band shares the same masks, which are based on W1, the deepest band. The first step of the masking process consists of finding the point sources brighter than W1$ = $15 in the cutouts and generating a preliminary mask with the {\tt sep mask\_ellipse} task \citep[python implementation of {\tt SExtractor}; see][]{barbary_sep_2018}. This task does not however allow for an efficient masking of the diffraction spikes associated with the brighter stars. We thus create a temporary mask by adding the star and unWISE masks together. Non-zero values in the unWISE mask correspond to bright stars, diffraction spikes, nearby galaxies, and saturated pixels. 

The unWISE coadds are background-subtracted. A constant is subtracted from W1 and W2 images, while a median filter is used to remove instrumental artifacts from the W3 images \citep{lang_unwise_2014}. We remove any residual gradient by masking every source in the science image (using the temporary mask for the bright sources and {\tt sep} for faint galaxies) and by fitting a plane (i.e.\ a polynomial of the first order) to the remaining pixels. This plane is then subtracted from the science image. Note that the gradient subtraction is performed independently in each band.

We then use the {\tt reproject} python package with the {\tt reproject\_exact}\footnote{See the documentation at \url{https://reproject.readthedocs.io/en/stable/\#module-reproject}} algorithm to merge eventual partial cutouts. Temporary masks are merged similarly.

We then further refine the merged masks in two steps. We first remove residual extended foreground galaxies, using the \citet{mancone_formation_2010} 3.4 $\mu$m luminosity function to interpolate the characteristic magnitude at the cluster redshift. Objects 2.5 magnitudes brighter than the characteristic magnitudes (i.e. ten times brighter than the characteristic luminosities) are then detected in W1 and masked in every band.

The final step of the masking process consists of masking non-photometric members. Since we have the photometric redshift PDF of each galaxy in the field of view, we adopt the following method. For each galaxy we compute the mode ($z_{mode}$) and standard deviation ($\sigma_z$) of the PDF. We then mask galaxies for which $| z_{mode} - z_{cluster} | > 3 \sigma_z$. The factor of three was determined empirically to minimize errors, while preserving the stack flux. 

W3 images are occasionally affected by improperly removed artifacts \citep{lang_unwise_2014}. To remove images significantly affected by bright artifacts, we sum the non-masked pixels of each image. We then use a sigma-clipping algorithm, with a 5$\sigma$ limit, to remove outliers within each bin. The discarded images correspond to about 1.3\% of the total W3 cutouts.

We weight each pixel separately in our stacks, using the sigma maps provided by the cutout service to compute the weight map. These sigma maps combine the errors computed by the WISE team with the standard deviation of the coadds \citep{lang_unwise_2014}. Wherever necessary, we add the partial sigma maps in quadrature to generate one sigma map per full cutout. Weight maps are then computed by squaring the inverse of the sigma maps. Each cutout and weight map pair are then rotated by a random multiple of 90\textdegree~before being stacked.

\subsubsection{GALEX image processing}\label{sssec_galex_pipeline}

One of the first challenges of GALEX image processing is the partial, patchy coverage. We deal with this issue in three steps. For each cluster, we test whether a single pointing (if any) will be enough to generate a full 15.9 arcmin$\times$15.9 arcmin cutout. In cases where this is not possible, we check whether any pointing covers a radius of 3 Mpc (which corresponds to the largest radius used in the analysis) around the cluster position, selecting the one with the longest exposure time if more than one pointing fulfills this condition. Finally, we can recover a few additional clusters by merging several pointings together. However, even with these adjustments, we are able to generate cutouts for only 46.8\% (i.e. 4849 clusters) of our total cluster sample.

The rest of the image processing is similar to WISE, with two notable adjustments. First, we use the $1546$ \AA~$L_*$ of \citet{moutard_uv_2020}, but caution that this monochromatic L$_*$ may be an overestimate at $z\gtrsim 1$, at which point the GALEX NUV filter stars to probe the Lyman break.

The second change compared to WISE is that the sigma maps must be computed. To do so, we assume that the Poisson noise is the dominant contribution to the noise. The conversion of the Poisson noise from counts to Janskys introduce a time dependency in the sigma map such as:

\begin{equation}\label{eq_sigma_GALEX}
    \sigma_{f_{Jy}} = \sqrt{\frac{A D f_{Jy}}{t}}
\end{equation}

\noindent where $f_{Jy}$ is the science pixel flux density before the background subtraction and D is the detector relative response, normalized such that the maximal response is equal to one. In this equation, $A$ is the conversion factor to transform the counts per second into Janskys (a constant) and $t$ is the exposure time in seconds.

\subsection{Herschel stacking method}\label{ssec_herschel_stacks}

\begin{figure}

    \centering

    \includegraphics[width=0.9\columnwidth]{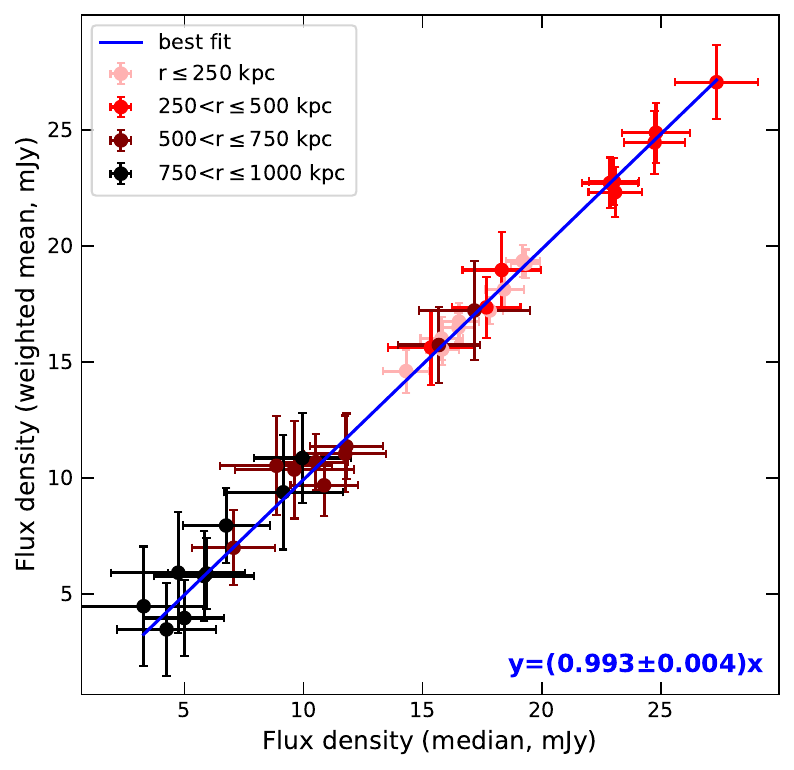}
    
    \caption{The relationship between the fluxes measured in a SPIRE 250 $\mu$m median stack and a weighted mean stack. The colors indicate different apertures.}
    \label{fig_median_wmean}
    
\end{figure}

Unlike WISE and GALEX which are divided into tiles or pointings, SPIRE and PACS images are presented in the form of already merged mosaics. We thus process each mosaic as a bloc and divide them into cutouts immediately before stacking. Masking is also considerably simplified, because SPIRE images are confusion-limited \citep{smith_herschel-atlas_2017,alberts_measuring_2021} and PACS images are shallow enough \citep[e.g.][]{smith_herschel-atlas_2017} that no individual cluster member is likely to be detected.

\subsubsection{SPIRE images}\label{sssec_spire_pipeline}

For SPIRE data, both raw and background-subtracted science maps are available. The raw maps display several bright extended regions due to the presence of Galactic cirrus, while the 1.5 arcmin scale used to compute the background in the latter set of maps (see Section \ref{ssec_herschel_data} and \citealt{valiante_herschel-atlas_2016}) is small enough that eventual diffuse emission associated with our clusters might have been removed. Since a 5 or 10 arcmin background scale does not efficiently remove the Galactic cirrus, we adopt the compromise of using the {\tt Background} task from {\tt sep} to locate and mask any regions where the background is at least
two standard deviations higher than the median background. We do not perform any further source masking; as explained in \citet{alberts_measuring_2021} the vast majority of cluster-related sources are unlikely to be individually detected, and the contributions of randomly placed foreground sources to the stack fluxes are minimal.

Confusion noise is one of the main sources of uncertainty in the SPIRE data and is thus an essential component of our sigma map. \citet{valiante_herschel-atlas_2016} gives an estimate of the mean confusion noise in each of the GAMA fields but does not include its variation with flux. Thus, we create ``confusion noise maps'' following instead the prescription of Table 6 of \citet{smith_herschel-atlas_2017} for the North and South Galactic Cap fields of the H-ATLAS survey. These maps are then added in quadrature to the HIPE-generated instrumental noise maps and divided into cutout matching the science images.

\subsubsection{PACS images}\label{sssec_pacs_pipeline}

\begin{figure*}[!ht]

    \centering
    \includegraphics[width=0.9\textwidth]{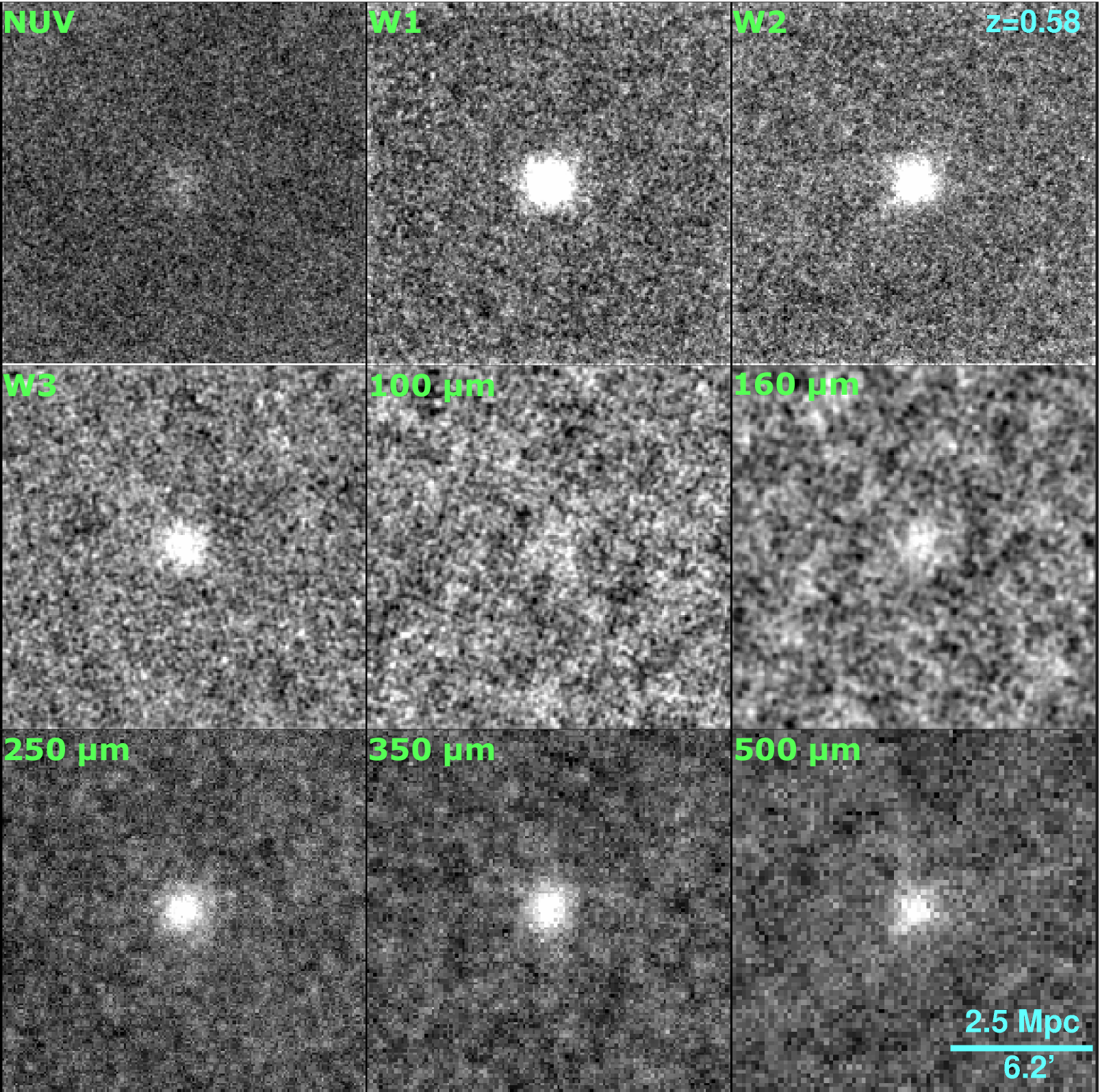}
    
    \caption{Stacks for $\overline{z} = 0.58$. The bands are indicated on the top left of each stack, and the approximate scale on the bottom right of the SPIRE 500 $\mu$m stack. Since the mean redshift of this bin is below 1.66, the W3 stack is not used in the analysis. The NUV, W3, 100, and 160 $\mu$m stacks have been smoothed by a Gaussian kernel with a size of 3 pixels and a $\sigma$ of 1.5 pixels to enhance the visibility of the central detection.}
    \label{fig_stack_0_575}
    
\end{figure*}

\begin{figure*}

    \centering
    \includegraphics[width=0.9\textwidth]{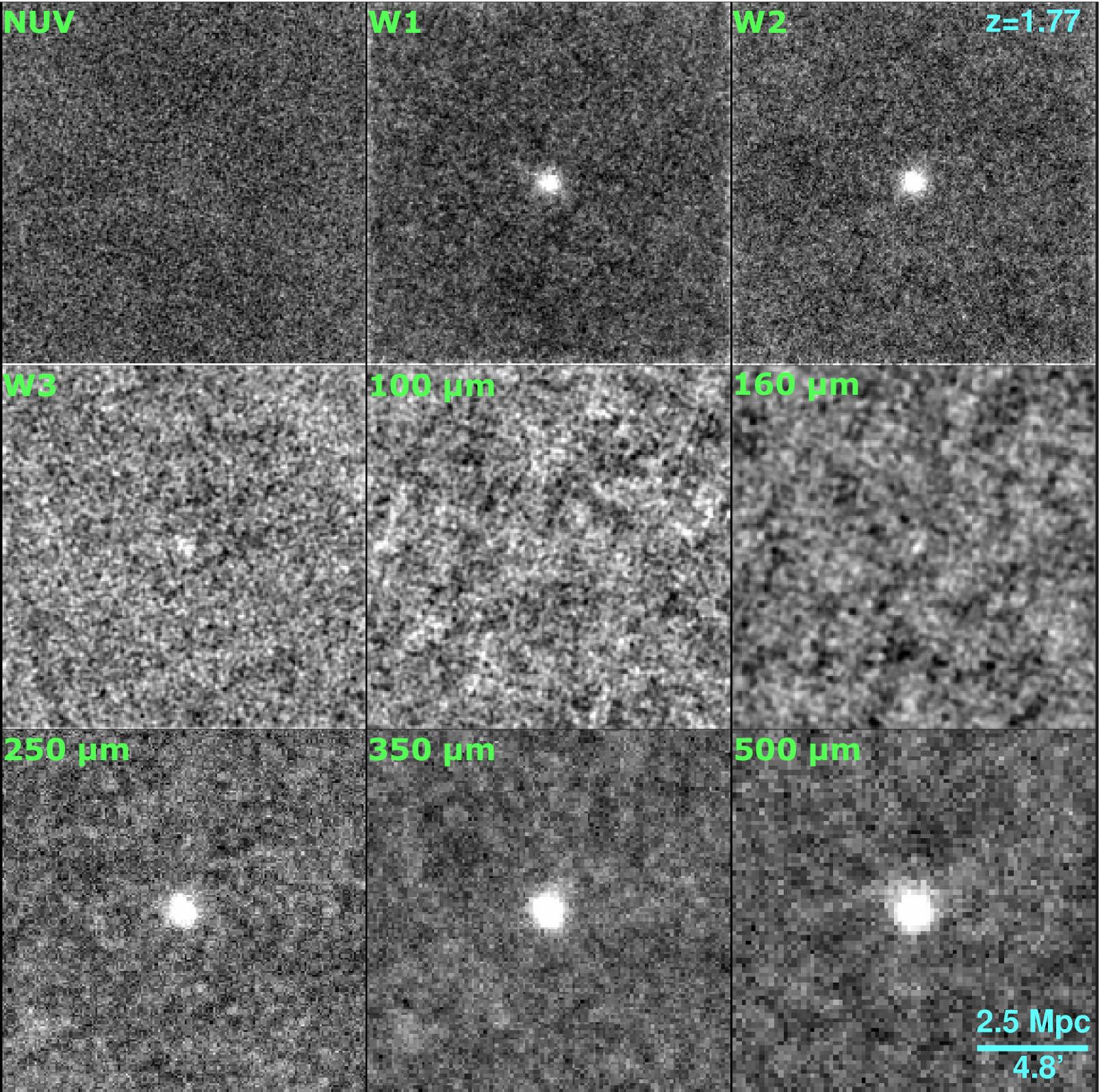}
    
    \caption{Stacks for $\overline{z} = 1.77$. The bands are indicated on the top left of each stack, and the approximate scale on the bottom right of the SPIRE 500 $\mu$m stack. The NUV, W3, 100, and 160 $\mu$m stacks have been smoothed by a Gaussian kernel with a side of 3 pixels and a $\sigma$ of 1.5 pixels to enhance the visibility of the central detection.}
    \label{fig_stack_1_772}
    
\end{figure*}

For PACS data, a background subtraction on a 4 arcmin scale was performed as part of the map-making process to remove large scale artifacts in the maps \citep{valiante_herschel-atlas_2016}. Thus, the dominant noise in PACS images is the instrumental noise, which is correlated with the noise of neighbouring pixels \citep[e.g.][]{ibar_h-atlas_2010,lutz_pacs_2011,popesso_effect_2012,valiante_herschel-atlas_2016}. According to Equation 9 in \citet{valiante_herschel-atlas_2016}, the 1$\sigma$ noise levels for a circular aperture with a radius of 4 pixels (12 and 16 arcsec) are 30.1 and 39.7 mJy in the 100 and 160 $\mu$m maps, respectively. In the regions where the number of scans is four (rather than two), these values are reduced by a $\mathrm{\sqrt{2}}$ factor.

We thus assume that every source in PACS is a foreground source. To mask these interlopers, we apply a sigma-clipping algorithm to our background-subtracted maps \---\ after multiplying them by the square root of the number of scan maps to remove the noise dependence on the number of scans. The upper bound of the algorithm is 3$\sigma$ and the lower bound is 10$\sigma$. 

The noise correlation represents a major challenge for the computation of a sigma map \citep{popesso_effect_2012}. We thus performed a median stacking of PACS cutouts, after rotating them by a random multiple of 90\textdegree~to circularize the PACS PSFs. To test the compatibility of the median stacks with the variance-weighted stacks used for other wavelengths, we generated median 250 $\mu$m stacks, which we compare to the variance-weighted ones. We then measure the flux densities in both set of stacks using the method presented in Section \ref{ssec_flux_ap_corr}, without the aperture correction. Results are shown in Figure \ref{fig_median_wmean}.

Figure \ref{fig_median_wmean} shows that the the slope of the relationship between both set of stacks is close to 1. Since both the PACS and SPIRE bands probe wavelengths dominated by dust emission, we expect the measurements obtained from PACS median stacks to be comparable with variance-weighted measurements. Furthermore, the deviation from a 1:1 relationship is negligible compared to the uncertainty associated with PACS fluxes.

\subsection{Bootstrap generation}\label{ssec_bootstrap}

We estimate the errors on the fluxes by generating a thousand bootstraps, with replacement. For example, if a stack contains 500 candidate clusters, we do 500 random selection of clusters among those same candidate clusters, with no restriction on the number of selections of a single candidate. We then rotate each frame by a random multiple of 90\textdegree, and stack.

For GALEX, we assemble the bootstraps using clusters for which we have coverage. If we have images for only 234 of the original 500 candidates, then each bootstrap will contain 234 frames, selected among the clusters with coverage. Similarly, we do not use the sigma-clipped frames in our W3 bootstraps. We find that for all wavelengths the flux distributions resulting from the bootstrap analysis are consistent with Gaussians with no significant outliers or large tails.

\begin{figure*}[!ht]

    \centering

    \includegraphics[width=0.9\textwidth]{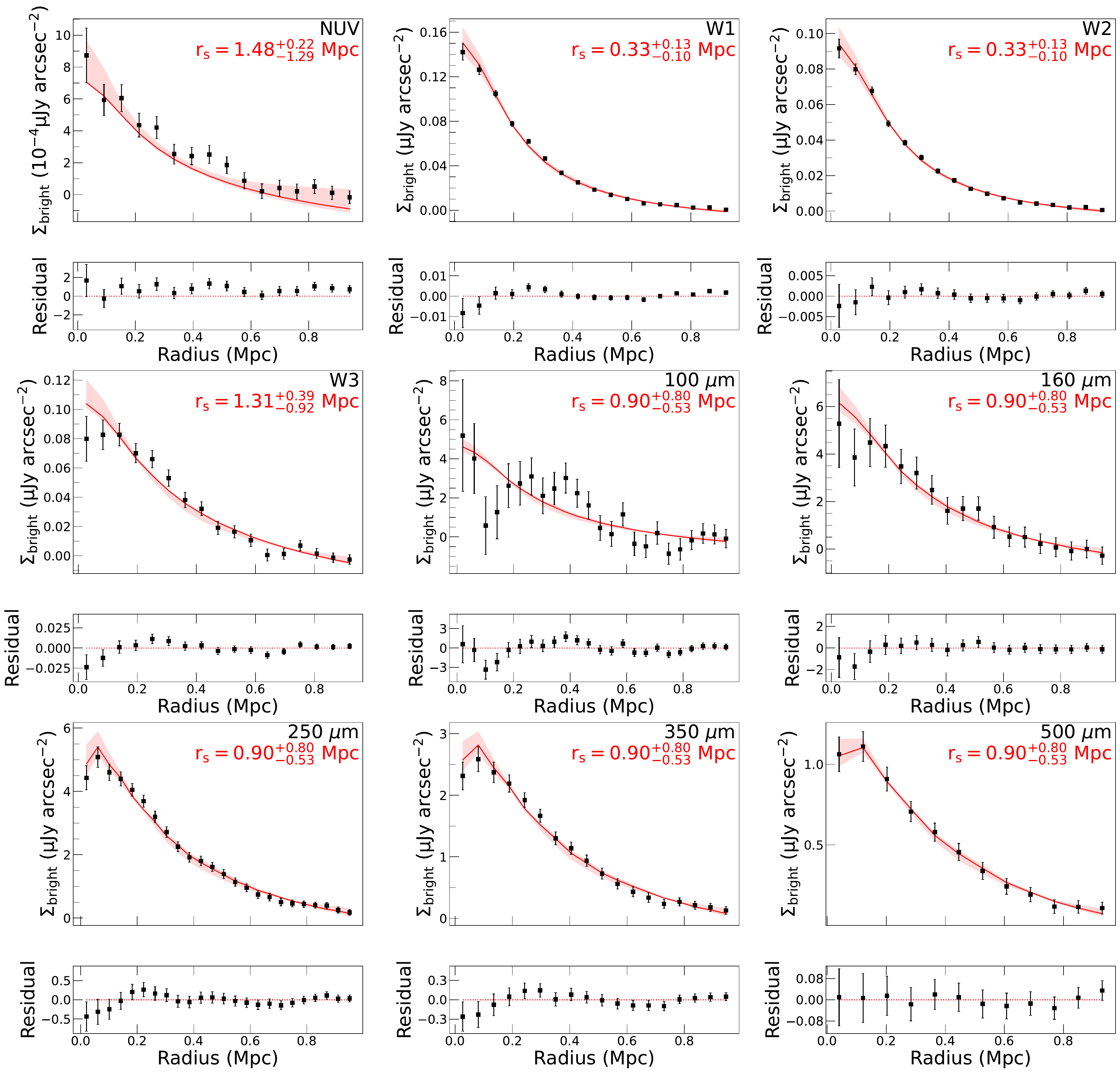}
    
    \caption{Differential surface brightness profiles of the $\overline{z} = 0.58$ stacks. The radii indicated on the \textit{x}-axis are the midpoint radii of each annular apertures. The best-fitting NFW profile, modelled to include the effect of miscentering and PSF blurring (see Section \ref{sssec_sim_profiles} explanation) is highlighted by a red curve. The shaded regions show  the range of posible surface brightnesses predicted by the models within the intersection of the prior with the 68\% confidence interval. The characteristic scale of the best-fitting NFW profile is indicated on the top right of each main panel. The smaller panels show the residual of the best fit.}
    
    \label{fig_profile_0_575}
    
\end{figure*}

\begin{figure*}[!ht]

    \centering

    \includegraphics[width=0.9\textwidth]{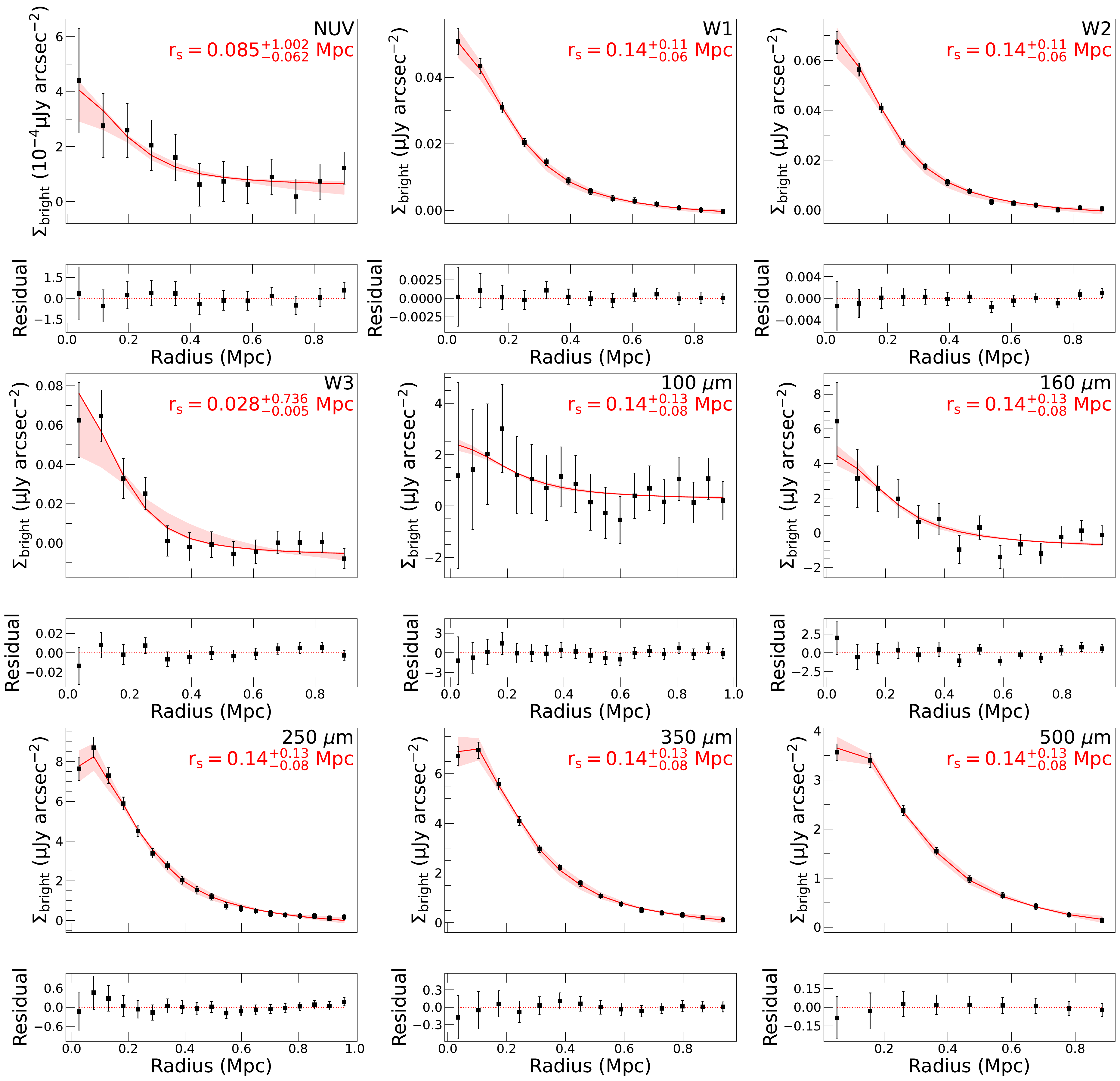}
    \caption{Same as Figure \ref{fig_profile_0_575}, but for the
    $\overline{z} =1.77$ stacks.}
    \label{fig_profile_1_772}
    
\end{figure*}

\section{Results}\label{sec_results}
Figures \ref{fig_stack_0_575} and \ref{fig_stack_1_772} present two examples of stacks, one at $\overline{z} = 0.58$ and one at $\overline{z} = 1.77$ respectively. Strong signals in W1, W2, and the SPIRE bands are present at all redshifts. W3 stacks, which sample the PAH at low redshift, are shown in both Figures but analysed only at $z \gtrsim 1.6$. GALEX clearly shows a central enhancement at $z \lesssim 1$ that fades at higher redshift, which is expected since at $z \gtrsim 1.77$ this band probes blueward of the Lyman break. When detected, GALEX fluxes provide constraints on the extinction and unobscured star formation. As {\tt CIGALE} SED fitting relies on an energy balance between the ultraviolet and the far infrared, the inclusion of GALEX data dramatically reduces the size of the error bars on the SFR. In the following sections, we examine the stack profiles and the steps involved in transforming these profiles into constraints on the SED.

\subsection{Radial profiles \& aperture correction}\label{ssec_aperture_correction}

The stacks presented above have PSF sizes ranging from a FWHM of about 5 arcsec (GALEX NUV) to about 36 arcsec (SPIRE 500 $\mu$m). Given this wide variety of resolutions, it is necessary to compute an aperture correction for each band. To do so, we adopt a forward modelling approach very similar to the one described in \citet{alberts_measuring_2021}.

We compare our stack profiles to NFW profiles, modelled to include the effects of miscentering and PSF blurring. An aperture correction is determined by comparing the flux densities of the blurred profile with an unaltered profile (i.e.\ without blurring and miscentering) with the same characteristic scale.

\subsubsection{Modelled profiles}\label{sssec_sim_profiles}

For each band and redshift bin in our analysis, we use the {\tt Colossus} package \citep{diemer_colossus_2018,diemer_accurate_2019} to compute the two-dimensional surface brightness of 80 NFW profiles: 40 well-centered profiles with characteristic scales ($r_s$) ranging from 0.015 to 1.9 Mpc in logarithmic steps; and 40 blurred and miscentered profiles with the same characteristic scales. To produce each of the 40 blurred profiles, we generate 100 NFW profiles with a random center. Each realization is convolved with the PSF of the band and then rotated by a random multiple of 90 degrees. They are then mean stacked to form a single blurred profile. To generate the random centers, we assume a Gaussian distribution with a standard deviation of 12 arcsec in x and y, based on the pixel scale of the density map used to detect the MaDCoWS2 clusters \citep{thongkham_massive_2024}. The PSFs are downloaded from the publicly available documentation\footnote{See \url{http://www.galex.caltech.edu/wiki/Public:Documentation/Chapter_106\#Point_Spread_Function} for GALEX, \url{https://wise2.ipac.caltech.edu/docs/release/allsky/expsup/sec4_4c.html\#coadd_psf} for WISE, and \url{https://www.h-atlas.org/public-data/download} 
for Herschel} except for the PACS bands, for which only the energy enclosed functions are available. We use these functions to reconstruct the PSFs assuming radial symmetry, which is reasonable given the random 90\textdegree~rotation applied to the cutouts in our stacks.

\subsubsection{Measured profiles}\label{sssec_obs_profiles}

To compute the aperture corrections, we determine which blurred NFW profile best matches the observed surface brightness. To do so, we measure the surface brightness in concentric annuli spaced by 9 and 8.25 arcsec in the GALEX and WISE images (6 and 3 pixels wide respectively). For Herschel data, we adapt our sampling to the varying resolution. We use 2 pixel wide annuli for the PACS data (6 and 8 arcsec for the 100 and 160 $\mu$m bands respectively) and 1 pixel wide annuli for the SPIRE data (6, 8 and 12 arcsec apart). This sampling is chosen to allow the SPIRE bands to be fitted together, while giving slightly more weight to the 250 $\mu$m band which has the best resolution.

When fitted individually, both the W1 and W2 have similar profiles, which is to be expected since they both probe the stellar component. To reduce the uncertainties associated with individual fits, we take the sum of the W1 and W2 $\chi^2$ and then determine the 68\% confidence interval of the combined $\chi^2$. GALEX NUV and W3 are fitted individually.

A similar reasoning should hold for the PACS and SPIRE bands, which all probe the warm dust content. While the strongly detected SPIRE band have indeed similar profiles, the PACS data provide at best marginal detections with rather uncertain profiles. We thus use the combined $\chi^2$ of the 250, 350, and 500 $\mu$m bands to constrain the far infrared NFW profiles, and apply those constraints on the PACS data as well.

To provide a measure of constraint on the aperture correction of the marginal detections, we implement a prior. Based on literature measurements \citep[see the compilation of][and references theirein]{alberts_clusters_2022}, we consider that a concentration \citep[$c = r_{200}/r_s$, see e.g.][]{navarro_structure_1996} of 1 to 10 is plausible. We thus calculate a prior corresponding to any characteristic scale between $M_{200} = 10^{13}~M_\odot$ and $M_{200} = 10^{15}~M_\odot$ assuming a concentration of 1 to 10. This corresponds to characteristic scales of $0.04 \leq r_s \leq 1.70$ Mpc at $\overline{z} = 0.58$ and $0.02 \leq r_s \leq 0.91$ Mpc at $\overline{z} = 2.37$ and to about 31 blurred NFW models within the prior per redshift. The 9 models outside the prior are used to estimate the limits of the intersection between the prior and the 68\% confidence interval with interpolation.

Figures \ref{fig_profile_0_575} and  \ref{fig_profile_1_772} show the stacks' surface brightness profiles, best-fitting models (combined for W1 and W2 and for the Herschel data; individual otherwise), and confidence intervals (within the prior) for $\overline{z} = 0.58$ and $\overline{z} = 1.77$.

Prior to profile fitting, we determined the average surface brightness of each stack between 2 and 3 Mpc, which we define as our background aperture, and subtract this background. This correction ensures that we remove the relative contribution of interloping galaxies from the stack and bring the flux to zero at large scales. We performed a similar operation on the bootstrap stacks used to compute the profile uncertainties. Without this correction SPIRE data tend to have surface brightnesses above zero at large radii, which is to be expected since the non-photometric members are not masked in these bands. In contrast, the uncorrected sky levels for the NUV and W3 data are systematically below zero. Since the vast majority of the emission in those bands comes from sources too faint to be individually detected, the background subtraction performed before the stacking of each frame could be systematically overestimated.

With these adjustments, the residual of the fits do not show systematic trends in the high redshift bins. However, at lower redshifts ($\overline{z} < 1.77$) the fitted profiles systematically overestimate the core surface brightnesses of the SPIRE and W3 profiles. \citet{popescu_tracing_2023} noted a similar overestimate in the center of their WISE stacks and suggested that an unaccounted offset between the image and the cluster centers might be the cause. Since we already correct for miscentering, this could indicate that the offset estimate incorporated in our blurred models might be slightly too optimistic. However, the absence of a noticeable systematic effect in the higher redshift profiles suggests that this effect is predominantly driven by the suppression of star formation in the cluster core at low and moderate redshifts \citep{alberts_measuring_2021}. We explore the evolution of the sSFR as a function of redshift and projected radius in Section \ref{sec_discussion}.

\subsection{Flux measurements and aperture corrections}\label{ssec_flux_ap_corr}

We measure the flux densities in our stacks in four concentric apertures: a circular aperture with a 250 kpc radius and then annular apertures between 250 and 500 kpc, 500 to 750 kpc, and finally 750 and 1000 kpc. We also use two additional circular apertures with radii of 1000 and 1500 kpc to perform checks and compare our results with the literature.

We correct the residual variations of the sky level in a similar fashion as in Section \ref{sssec_obs_profiles}: we measure the surface brightness in an annular aperture between 2 and 3 Mpc, multiply it by the area of each science aperture and subtract the result from each flux density. This operation is made both in our science and bootstrap stacks to take into account the impact of this subtraction in our flux errors.

To test the robustness of this correction to an increased fraction of interlopers, we tried more lenient versions of non-members masking, with first a tolerance to deviation of up to 4$\sigma$ from the cluster redshift (see Section \ref{ssec_wise_galex_stacking}). We tested also an extreme case of 10$\sigma$ tolerance (i.e. basically no masking of the non-members). In both cases, the change in the sSFR profile are much smaller than the size of the error bars.

Aperture corrections are determined by measuring the flux in the unblurred, well-centered NFW images and dividing by the flux from the corresponding blurred images. We determine the intersection of the prior and the 68\% confidence interval to estimate the uncertainty on the fluxes, using linear interpolation to calculate the aperture corrections at the edges of the interval. The aperture-corrected fluxes are given in Appendix~\ref{sec_table_fluxes}.

Typical aperture corrections for W1 and W2 are in the 1.10 to 1.20 range  for the innermost aperture, and between 0.89 and 0.97 for the apertures between 250 and 1000 kpc of projected radius. SPIRE 250 $\mu$m aperture corrections are between 1.15 and 1.41, and 0.88 and 1.01, for the core and the other apertures respectively, while SPIRE 500 $\mu$m aperture corrections fall between 1.2 and 1.60 for the core, and between 0.88 and 0.97 for the other apertures. As a sanity check, we computed our sSFRs with and without an aperture correction. The absence of an aperture correction decreases the stellar masses and the star formation rates in the core, and slightly increases them between 250 and 500 kpc. However, the two effects compensate each other so the sSFR is nearly unchanged.

\begin{figure}

    \centering

    \includegraphics[width=0.9\columnwidth]{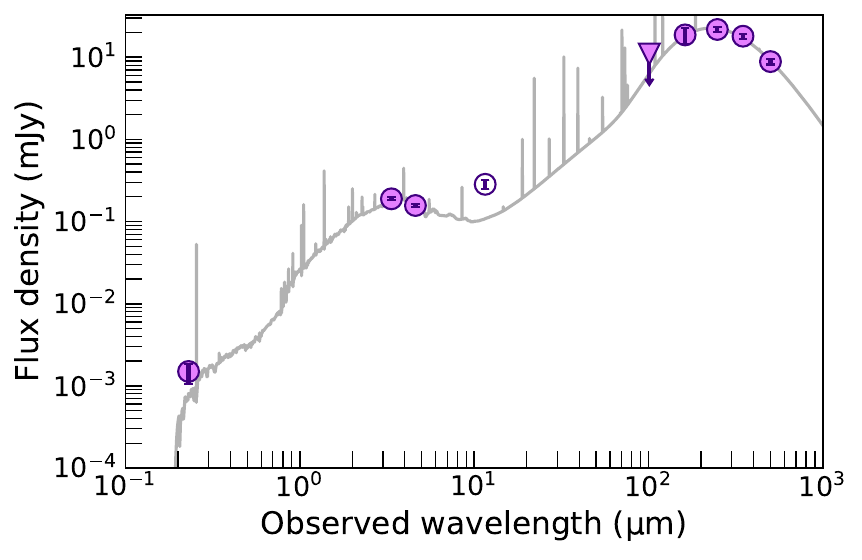}
    
    \caption{Comparison between {\tt CIGALE} best-fitting SED (in grey) and the flux densities measured between 250 and 500 kpc at $\overline{z} = 1.1$. A downward triangle denotes the upper limit at 100 $\mu$m. W3, which is not used in this fit because it probes PAH emission, is indicated by an open symbol. The best fit from {\tt CIGALE} corresponds to an sSFR of 0.25 Gyr$^{-1}$ and has a characteristic time of 2 Gyrs, with the age of the oldest stars being 5 Gyrs. The dust temperature is 45K and the E(B-V) attenuation is 0.3 magnitudes. {\tt CIGALE}'s reduced $\chi_\nu^2 = 1.01$.}
    
    \label{fig_SED_z_1_101_500kpc}
    
\end{figure}

\subsection{SED fitting}\label{ssec_SED}

\begin{figure*}

    \centering

    \includegraphics[width=0.9\textwidth]{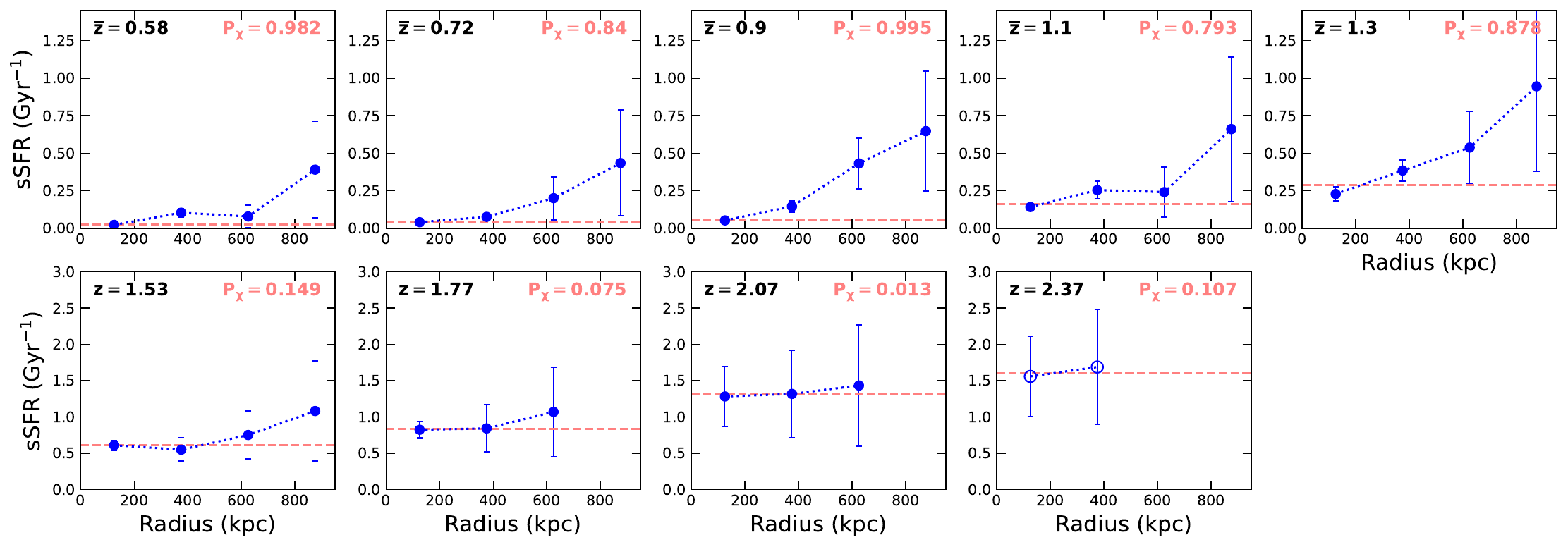}

    \caption{Specific star formation rate as a function of the aperture. Each aperture is represented by the projected radii corresponding to its midpoint (e.g. 125 kpc for the aperture between 0 and 250 kpc). We test the hypothesis of a null slope (in pink) and find that $\overline{z} \lesssim 1.5$ are inconsistent with a constant sSFR. The $P_\chi$, the probability of obtaining a lower $\chi^2$ with a random set of data from the parent distribution, is indicated at the top right of each panel. The thin black line at 1 Gyr$^{-1}$ was added to facilitate comparison between panels.}
    
    \label{fig_sSFR_vs_ap}
    
\end{figure*}

We use {\tt CIGALE} \citep{burgarella_star_2005,noll_analysis_2009,boquien_cigale_2019} to calculate the SFRs and the stellar masses of our stacks. {\tt CIGALE} assumes energy conservation between the dust absorption and emission. The code calculates the $\chi^2$ likelihood for a fixed grid of models and returns the likelihood-weighted mean and standard deviation of each parameter \citep[see Section 4.3 of][]{boquien_cigale_2019}.

{\tt CIGALE} includes three options for star formation histories: $\tau$-model, delayed $\tau$-model and periodic. We adopt the delayed $\tau$-model because it has been reported in the literature \citep[e.g.][]{maraston_star_2010,pacifici_importance_2015,carnall_how_2019} that the $\tau$-model cannot reproduce the rise and fall of the SFR, which can be a problem when modelling star-forming or recently quenched galaxies \citep{pforr_recovering_2012,boquien_cigale_2019}.

We assume an attenuation law based on \citet{calzetti_dust_2000} and \citet[][module \textsc{dustatt\_calzleit} in {\tt CIGALE}]{leitherer_global_2002}, and a \citet{casey_far-infrared_2012} dust emission model with an emissivity index $\beta = 1.6$ and a mid-infrared power law index $\alpha = 2$. We use \citet{bruzual_stellar_2003} simple stellar population models, assuming solar metallicity. We also use the \textsc{nebular} module \citep{inoue_rest-frame_2011} to model the emission lines associated with nebular emission. The free parameters and their possible values are listed in Table \ref{tab_cigale}. Unless otherwise stated, every other parameter is fixed at default values.

\begin{deluxetable}{lcc}
\tablecaption{Number of clusters in each redshift bin}\label{tab_cigale}
\tablehead{
\colhead{Parameter} & \colhead{Range} & \colhead{Step}
}
\startdata
Age (Myr)$^a$ & 250 to 8500 & 250\\
Characteristic time (Myr) & 250 to 8500 & 250\\
Dust temperature (K) & 15 to 70 & 5\\
$E(B-V)$ (mag) & 0 to 1 & 0.1\\
\enddata
$^a$ Values larger than the age of the Universe at the considered redshift are rejected by CIGALE.
\end{deluxetable}

It should be noted that the \citet{casey_far-infrared_2012} model does not attempt to reproduce emission from the PAH features. While other {\tt CIGALE} modules model the PAH emission, it is customary for most stacking analysis \citep[e.g.][]{mckinney_measuring_2022,popescu_tracing_2023} to ignore the flux of the bands that overlap with the PAH region. \citet{alberts_clusters_2022} demonstrated that within a range of $\Delta z = 0.3$, the W3, W4 and Spitzer 24 $\mu$m emission of individual clusters can vary by factor of 1.2 to 2.5 due to the sole redshift offsets. This additional scatter is not expected by most SED fitting codes, and can lead to discrepant results if the PAH are included.

The aperture-correction introduces a small asymmetry between our upper and lower uncertainties (usually less than a 20\% difference), which cannot be handled properly by {\tt CIGALE}. We thus symmetrize our error bars by taking the larger value. We use upper limits for $< 2\sigma$ detections. To properly constrain the stellar mass and SFR, we remove every aperture where both W1 and W2 are upper limits or where the three SPIRE data points are upper limits. Figure \ref{fig_SED_z_1_101_500kpc} shows a comparison between the data for a 250 to 500 kpc aperture on the $\overline{z} = 1.1$ stacks and the best-fitting SED model generated by {\tt CIGALE}. The 100 $\mu$m upper limit is indicated by an arrow.

\section{Discussion}\label{sec_discussion}

\subsection{The onset of environmental quenching in the cluster core}\label{ssec_ssfr_vs_profile}

Figure \ref{fig_sSFR_vs_ap} presents the evolution of the sSFR as a function of the projected radius and redshift. At $\overline{z} < 1.3$, the sSFR tends to increase with the projected radius. Between $\overline{z} = 1.3$ and $\overline{z} = 1.77$, the central decrement progressively disappears, until the sSFRs at different radii become comparable. However, there is a marked increase of the modelling uncertainty at $\overline{z} \geq 1.77$ due to the poor constraints on the UV flux.

While the uncertainties are too large to robustly measure a change in the shape of the decrement, we can test the robustness with which a decrement is detected. To do so, we test the null hypothesis of no radial trend. We find that the profile are consistent with this null hypothesis only at $\overline{z} \geq 1.53$.

\begin{figure*}

    \centering
    \includegraphics[width=0.9\textwidth]{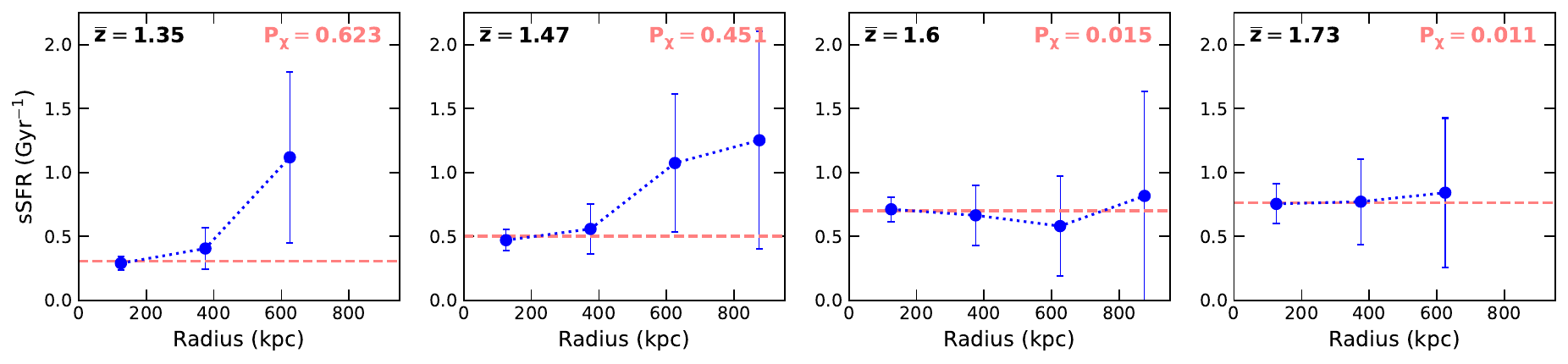}
    
    \caption{Specific star formation rate as a function of the stacks projected radii for a subset of redshifts with a tighter binning. The best-fitting horizontal line is indicated in pink, with the associated $P_\chi$ at the top right of each panel. The flattening of the profile between $\overline{z} = 1.35$ and $\overline{z} = 1.60$ is more obvious than in Figure \ref{fig_sSFR_vs_ap}, but the large modelling uncertainties mean that the profile is moderately consistent with the null hypothesis in every panel.}
    \label{fig_sSFR_vs_ap_subset}
    
\end{figure*}

To further investigate the flattening of the sSFR profile, we use a narrower bin definition for a subsample of 2450 clusters at $1.3 < z < 1.8$ (See Section \ref{ssec_madcows2} and the bottom panel of Figure \ref{fig_binning}). The resulting sSFR profiles, presented in Figure \ref{fig_sSFR_vs_ap_subset}, show a clearer change in the profile, happening between $\overline{z} = 1.35$ and $\overline{z} = 1.60$.

\begin{figure*}

    \centering

    \includegraphics[width=0.9\textwidth]{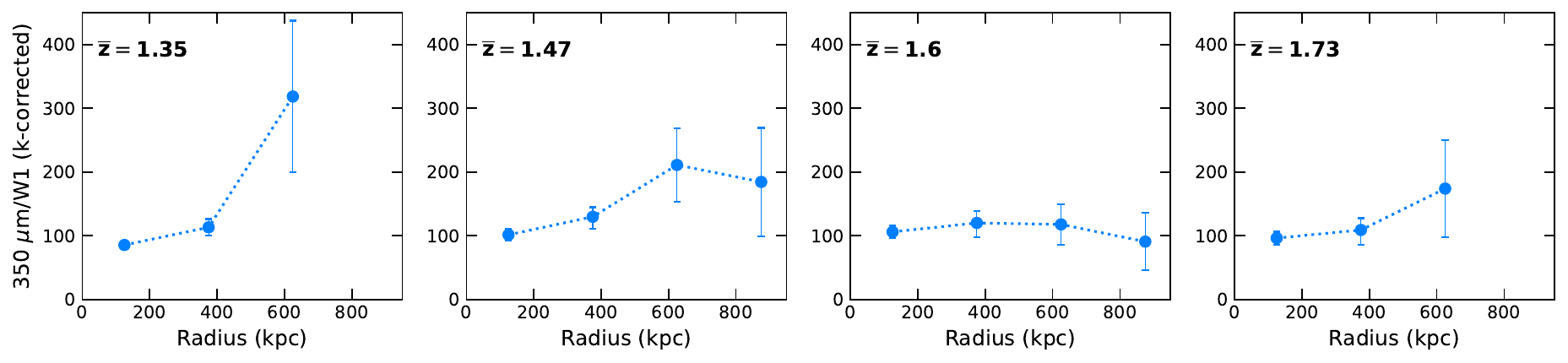}
    
    \caption{The evolution with redshift of the 350 $\mu$m/W1 ratios (k-corrected to $z = 1$) as a function of the projected physical radii. This ratio evolution provides an additional check on the profile evolution since bands W1 and 350 $\mu$m trace the stellar mass and the obscured SFR at $z = 1$ respectively, without the modelling uncertainties.}
    \label{fig_kcorr_fluxes_subset}
    
\end{figure*}

However, given the large modeling errors introduced by the {\tt CIGALE} fits, we perform an additional check, using the 350 $\mu$m/W1 ratios (k-corrected to $z = 1$) as a proxy for the sSFR. Specifically, to keep the k-correction to a minimum, we measure the fluxes in the bands closest in wavelength to the portion of the SED probed by W1 and SPIRE 350 $\mu$m at $z \sim 1$. We then use the best-fitting SED model provided by our {\tt CIGALE} fit to compute  k-corrections, which vary between -0.17 and 0.13 for W1, and between -0.31 and 0.24 for SPIRE 350 $\mu$m.

The resulting profiles, presented in Figure \ref{fig_kcorr_fluxes_subset}, demonstrate that the trend observed in Figure \ref{fig_sSFR_vs_ap_subset} is data-driven, with the additional modelling uncertainties originating mostly from the lack of UV constraints at $z \gtrsim 1.6$. The flat profiles observed at this epoch suggest that at high-redshift the local density does not have an impact on the star formation activity, i.e.\ that environmental quenching is unimportant at $z \gtrsim 1.6$.

The elapsed time is about $\sim$ 630 Myrs between $\overline{z} = 1.35$ and $\overline{z} = 1.60$. This suggests that the main mechanism responsible for the sSFR decrease in the core operates on a fast timescale \---\ which tentatively favors ram-pressure stripping \citep{gunn_infall_1972,poggianti_star_1999,lotz_gone_2019} or overconsumption \citep{brodwin_era_2013,mcgee_overconsumption_2014,balogh_evidence_2016} over slower mechanisms such as starvation \citep{larson_evolution_1980,balogh_origin_2000} or harassment.

We cannot completely rule out quenching by starvation: if one assumes a delayed-then-rapid quenching model, the fading time (i.e. the time to quench the galaxy after the delay) is about 0.4 and 0.8 Gyr for $\gtrsim 10^{10}~M_\odot$ galaxies \citep{wetzel_galaxy_2013,balogh_evidence_2016,fossati_galaxy_2017}. However, \citet{balogh_evidence_2016} and \citet{fossati_galaxy_2017} find that the delay time increases with decreasing stellar mass, with a delay time of about 5 Gyr at z $\sim 1$ for the less massive galaxies in their samples \---\ longer than the age of the Universe at z $\sim 1.5$. Thus, the delay time would need to be significantly shorter at high redshift for starvation to explain the change in our profiles.

Our sSFR profiles are broadly consistent with previous studies,
which suggests a transition between a profile where the star-formation activity increases with the projected radius to a relatively flat \citep{brodwin_era_2013,alberts_star_2016} or even decreasing profile \citep{tran_reversal_2010,santos_reversal_2015} at high redshift. The work of \citet{brodwin_era_2013} and \citet{alberts_star_2016} suggest a transition time around $z \sim 1.4$, which is slightly later, but still consistent with what we measure in our sample. 
We note however, that these studies might not be comparable to ours: they rely on small samples (16 and 11 clusters respectively) of clusters with a narrow mass range. Furthermore, both studies focus on massive galaxies \---\ stellar masses higher than $\mathrm{1.3\times10^{10}~M_\odot}$ \---\ which might behave differently than the more general population of cluster members. 

\subsection{Comparison with the field evolution}\label{ssec_ssfr_z_field}

\begin{figure*}

    \centering

    \includegraphics[width=0.9\textwidth]{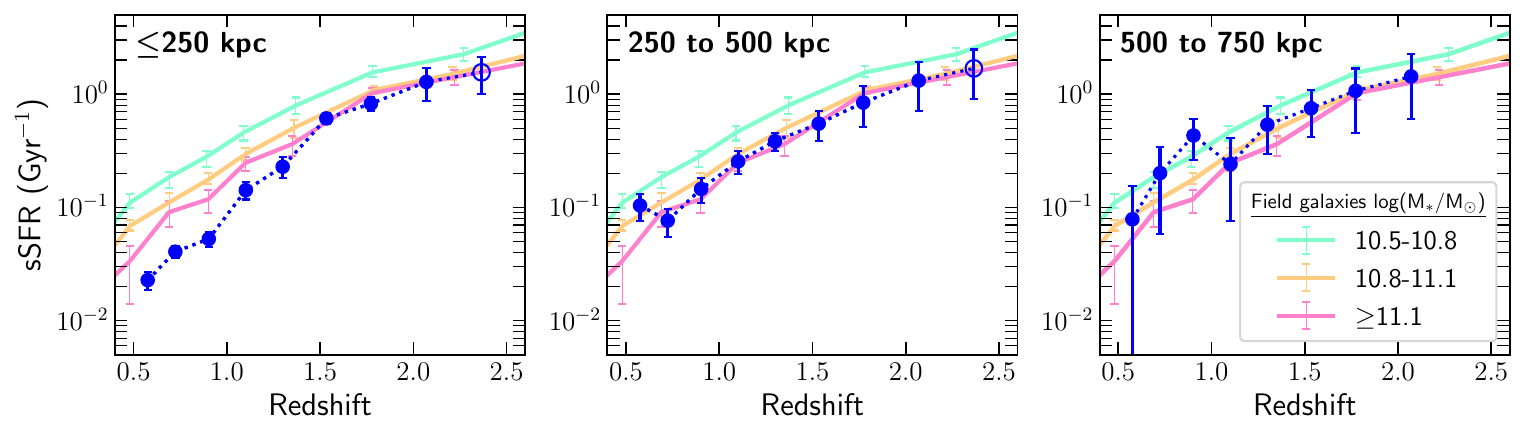}

    \caption{The redshift evolution of the sSFR in the three innermost apertures (blue dots), compared with the field evolution for different galaxy stellar masses. Field sSFR measurements are drawn from \citet{karim_star_2011}. We note that below $\overline{z} = 1.53$, our sSFR measurements in the core drop faster than in the field.}
    
    \label{fig_sSFR_vs_z}
    
\end{figure*}

\begin{figure*}

    \centering

    \includegraphics[width=0.9\textwidth]{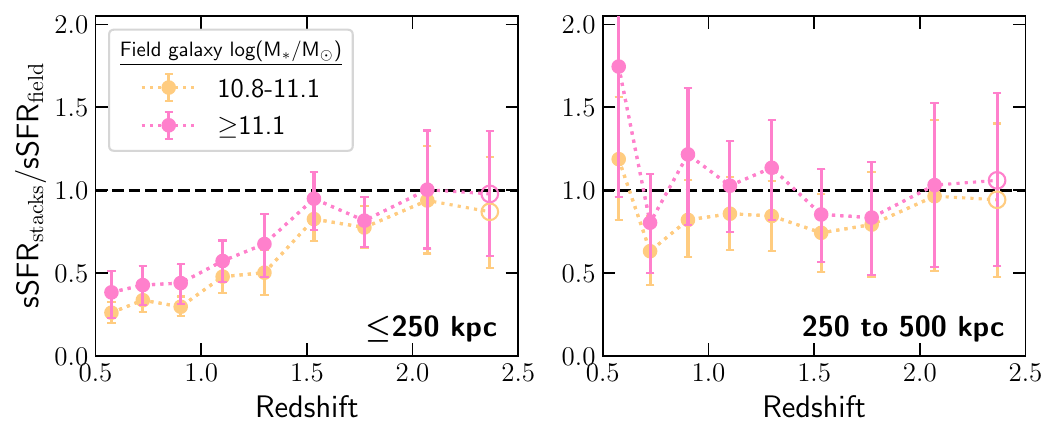}
    \caption{Cluster sSFRs normalized by the field sSFRs for the two more massive bins in \citet{karim_star_2011}. The sSFR in the cluster cores falls below the field sSFR at $\overline{z} = 1.53$.}
    \label{fig_sSFR_divided_field}
    
\end{figure*}

While our results agree with the literature on the specific topic of cluster profiles, as noted in Section \ref{sec_intro} the general literature surrounding the importance of quenching at $z \gtrsim 1.5$ is less clear, with reports pointing both toward and against environmental quenching being important at this epoch.

The most commonly used method to determine whether environmental quenching operates at high redshifts is to measure the fraction of red, passive-looking galaxies in a sample of clusters, compared with a similar measurement in the field. While the light-integrated nature of our stack is not compatible with the calculation of a quenched fraction, we can still perform a comparison with the field galaxy evolution.

To do so, we compare the sSFRs measured in our three innermost apertures to the field sSFR evolution, as traced by stacked radio emission in \citet{karim_star_2011}. Figure \ref{fig_sSFR_vs_z} shows the results of this comparison for three different field galaxy masses and three apertures.  In the innermost aperture, the sSFRs we measure are consistent with the field values for galaxies with $\log(M_*/M_\odot)\geq 10.8$ at $z \gtrsim 1.5$, but fall below the field sSFRs at lower redshifts. This is highly suggestive of a transition in the core between an evolution dominated by internal quenching processes to an evolution where environmental processes play a significant role. Interestingly, the form of the evolution of our sSFRs beyond a projected radius of 250 kpc appears consistent with field evolution at all the probed redshifts. It is thus unclear whether the high quenched fractions observed in the outskirts of local clusters result mostly from environmental quenching in the main halo \citep[e.g.][]{zinger_quenching_2018,brambila_examining_2023,kim_gradual_2023} or from pre-processing \citep[e.g.][]{werner_satellite_2022,lopes_role_2024}.

Figure \ref{fig_sSFR_divided_field} shows a different view of the field-cluster comparison presented in Figure \ref{fig_sSFR_vs_z}: the ratio of our measurements relative to the field sSFRs for the two highest mass bins from \citet{karim_star_2011}. Below $z \lesssim 1.6$, the sSFR in the cluster core falls steadily relative to the field sSFR while at higher redshifts the cluster core sSFR and the field sSFR are indistinguishable. At larger cluster radii there is no clear difference between field and cluster sSFRs at any redshift. 

\begin{figure}

    \centering
    \includegraphics[width=0.9\columnwidth]{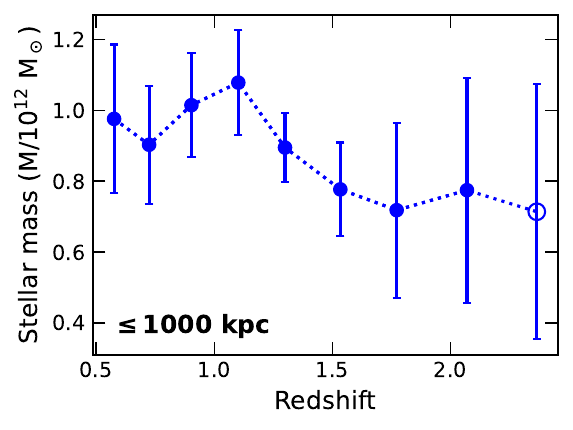}
    
    \caption{The redshift evolution of the total stellar mass enclosed within 1 Mpc of the stack centers.}
    \label{fig_total_stellar_mass}
    
\end{figure}

As a test of the robustness of our sSFR measurements, we measure the stellar mass in  a circular aperture with a 1000 kpc radii \---\ large enough to contain most of the stacked light. Figure \ref{fig_total_stellar_mass} presents the results. The stellar mass changes little with redshift, with perhaps a small increase toward lower redshift. Thus, the decrease of the sSFR in the core is unlikely to be solely due to a change in the total stellar masses across redshift bins. We repeat this test with a 1.5 Mpc aperture as a consistency check. The results are consistent, albeit with significantly larger uncertainties than for the 1 Mpc aperture. We are thus confident that the residual background corrections applied in Section \ref{ssec_aperture_correction} (which are based on the surface brightness between 2 and 3 Mpc) do not result in underestimated sSFRs. Either a bigger sample or one with more, or deeper, bands in the near-infrared will be needed to provide further constraints on the stellar mass and thus on the sSFRs beyond 1000 kpc. The Euclid Observatory could provide better constraints on the stellar mass evolution, but its data will need to be complemented by observations in the far infrared and UV to provide a complete picture of the star formation history in galaxy clusters and groups.

\subsection{The dependence of the sSFR on the S\slash N}\label{ssec_ssfr_vs_SNR}

\begin{figure*}

    \centering

    \includegraphics[width=0.9\textwidth]
    {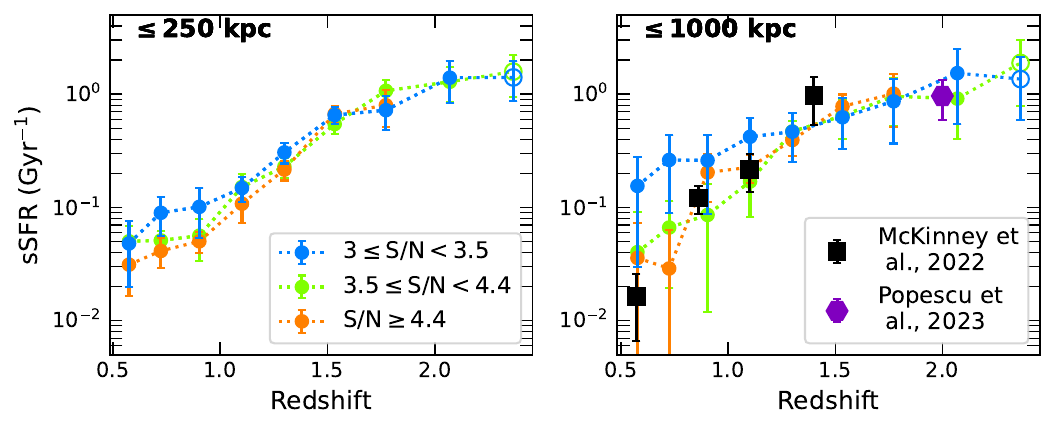}
    
    \caption{Comparison between the redshift evolution of the sSFR in three of our S\slash N subsamples, and with the \citet{mckinney_measuring_2022} sample. At $z \lesssim 1.2$, we note a modest difference in sSFRs between the highest and lowest S\slash N bins. The \citet{mckinney_measuring_2022} sample evolution appears globally consistent with the S\slash N$\sim$4 and 5 bins. At high redshift, the \citet{popescu_tracing_2023} $z \sim 2$ protocluster measurement is consistent with our data, despite the larger aperture and different sample properties.}
    \label{fig_McKinney_comparison}
    
\end{figure*}

To explore the dependence of the sSFR on the cluster masses, we subdivide our sample into S\slash N bins, as described in Section \ref{ssec_madcows2}. Figure \ref{fig_McKinney_comparison} shows the sSFR evolution with redshift in three S\slash N bins \---\ we omit the highest S\slash N bin since it does not span a large range of redshifts \---\ compared with sSFR based on \citet{mckinney_measuring_2022} work. We also include an sSFR based on the lower redshift bin of the \citet{popescu_tracing_2023} Planck-selected protoclusters \citep{planck_collaboration_planck_2015}, although the aperture used for their flux measurements is significantly larger than ours (7 arcmin \---\ about 3.6 Mpc at $z \sim 2$).

\citet{mckinney_measuring_2022} add UV constraints to the stacking analysis of \citet{alberts_measuring_2021}. Their sample covers the $0.5 < z_{phot} < 1.6$ redshift range with 232 clusters from the Infrared Array Camera (IRAC) Shallow Cluster Survey \citep[ISCS; ][]{eisenhardt_clusters_2008}. The median halo mass of their sample is $\log(M_{200}/M_\odot) = $13.7-13.9 \citep{alberts_evolution_2014} and has no significant evolution with redshift. Based on \citet{thongkham_massive_2024} scaling relation, we expect these halo masses to correspond roughly to S\slash N of 3.5-6 at $z \sim 0.5$ and to S\slash N of 3-4.5 at $z \sim 1.6$. \citet{mckinney_measuring_2022} assume an initial period of star formation modelled by a $\tau$-model followed by a burst at later time. Since this is significantly different from our own assumptions, we reprocess their flux density measurements, using the {\tt CIGALE} settings described in Section \ref{ssec_SED}. We similarly reprocess the \citet{popescu_tracing_2023} fluxes.

The sSFRs of the different S\slash N bins in Figure \ref{fig_McKinney_comparison} are indistinguishable from one another above $\overline{z} = 1.3$. Below this redshift, there is tentative evidence that the sSFR falls less rapidly for the lowest mass clusters (lowest S\slash N), both in the core and in the cluster overall. The \citet{mckinney_measuring_2022} cluster sample sSFRs are consistent with our measurements for the two higher S\slash N bins. We stress however that the relationship between the S\slash N and the total cluster mass is not calibrated at $z \gtrsim 1.2$ and might suffer from significant scatter \citep{thongkham_massive_2024}.

\section{Conclusions}\label{sec_conclusion}
 
We have performed a stacking analysis of 10,353 galaxy clusters and large groups from the MaDCoWS2 survey. Our work builds on the techniques developed by \citet{alberts_measuring_2021} and \citet{mckinney_measuring_2022} to analyze the integrated galaxy population of a sample of galaxy clusters, but focuses on the evolution of the sSFR with redshift and projected radii. We measure the sSFR in a central circular aperture (250 kpc radius), and then in three annular apertures at 250 to 1000 kpc. Our main findings can be summarized as follow:

\begin{enumerate}
    \item The sSFR tends to increase with projected cluster radius up to a redshift of $\overline{z} = 1.35$. At $\overline{z} \gtrsim 1.60$, the sSFR varies little with radius. This flattening of the profile is consistent with the literature profiles inferred from massive cluster members \citep{brodwin_era_2013,alberts_star_2016}.

    \item We interpret this transition as the onset (or dramatic increase) of environmental quenching in the cluster core. The relatively short time span over which the transition occurs (about 630 Myr between $\overline{z} = 1.35$ and $\overline{z} = 1.60$) implies that the dominant environmental quenching process might operate on a short timescale. This is tentative evidence that ram pressure stripping or overconsumption might be more important than starvation or galaxy harassment in this redshift regime.

    \item A comparison with \citet{karim_star_2011} indicates that the sSFR in the cluster cores is consistent with field sSFR levels for massive galaxies for $\overline{z} > 1.53$. At lower redshifts, while both are declining, the sSFR in cluster cores falls steadily below the field sSFR. This is further evidence that environmental quenching starts to operate in the core below a redshift of 1.5. Our data are inconclusive about an eventual influence of the environment in the outskirts \---\ at projected radii larger than 250 kpc there is no evidence of an accelerated drop compared to the field sSFR levels.
    \item Our results present tentative evidence that sSFR decreases with cluster mass, but the effect is modest at best. Larger samples of clusters will be needed to pinpoint when the mass-dependent quenching in clusters began.
    
\end{enumerate}
In a context where the importance of environmental effects at high redshift is widely debated, these results provide statistical evidence that environmental quenching in groups and clusters is not important at $z \gtrsim 1.5$. Although some cluster members may be environmentally quenched earlier, and some clusters may experience widespread environmental quenching in their cores before $z \sim 1.5$, our results suggest that such systems are not representative of the general population of cluster members.


\begin{acknowledgments}

We thank Dustin Lang for making WISE mask cutouts available on the Legacy viewer website. This material is based upon work supported by the National Science Foundation under Grant No. 2008367. The authors acknowledge University of Florida Research Computing for providing computational resources and support that have contributed to the research results reported in this publication. KSL and EP acknowledge financial support from the National Science Foundation under grant No. AST-2206705 and from NASA through the Astrophysics Data Analysis Program, grant number 80NSSC19K0582. The work of P.E. was carried out at the Jet Propulsion Laboratory, California Institute of Technology, under a contract with the National Aeronautics and Space Administration (80NM0018D0004). The unWISE coadded images and catalog are based on data products from the Wide-field Infrared Survey Explorer, which is a joint project of the University of California, Los Angeles, and the Jet Propulsion Laboratory/California Institute of Technology, and NEOWISE, which is a project of the Jet Propulsion Laboratory/California Institute of Technology. WISE and NEOWISE are funded by the National Aeronautics and Space Administration. The Herschel-ATLAS is a project with Herschel, which is an ESA space observatory with science instruments provided by European-led Principal Investigator consortia and with important participation from NASA. The H-ATLAS website is http://www.h-atlas.org/. GALEX is a NASA Small Explorer, and we gratefully acknowledge NASA’s support for construction, operation, and science analysis for the GALEX mission, developed in cooperation with the Centre National d’Etudes Spatiales of France and the Korean Ministry of Science and Technology. This research has made use of the Spanish Virtual Observatory (https://svo.cab.inta-csic.es) project funded by MCIN/AEI/10.13039/501100011033/ through grant PID2020-112949GB-I00.

\end{acknowledgments}

\vspace{5mm}
\facilities{GALEX, Herschel (PACS, SPIRE), WISE}

\software{Astropy \citep{astropy_collaboration_astropy_2013,astropy_collaboration_astropy_2018,astropy_collaboration_astropy_2022},
Astroquery \citep{ginsburg_astroquery_2019},
Cigale \citep{boquien_cigale_2019},
Colossus \citep{diemer_colossus_2018,diemer_accurate_2019},
Matplotlib \citep{hunter_matplotlib_2007}, 
NumPy \citep{harris_array_2020},
Reproject \citep{robitaille_reproject_2020,robitaille_astropyreproject_2023},
SciPy \citep{virtanen_scipy_2020},
Sep \citep{barbary_sep_2018}
Spanish Virtual Observatory \citep[online][]{Rodrigo_SVO_2012,Rodrigo_SVO_2020}}

\appendix

\section{Aperture-corrected fluxes} \label{sec_table_fluxes}

Table \ref{tab_fluxe} presents the aperture corrected flux of the sample, with no S\slash N subdivisions. Fluxes of the form ``<X'' indicate 2$\sigma$ upper limits. For $\overline{z} \lesssim 1.66$, WISE3 measurements are probing the PAH region and are not used for the SED fitting. Aperture-corrected WISE3 fluxes are given here for illustrative purposes only.

\begin{deluxetable*}{llccccccccc}
\tablecaption{Aperture-corrected flux densities and 2$\sigma$ upper limits for the main stacks. At $\overline{z} \lesssim 1.66$, WISE3 measurements probe the PAH and are thus not used in the SED analysis. They are given here for illustrative purposes only.}\label{tab_fluxe}
\tablewidth{700pt}
\tabletypesize{\scriptsize}
\tablehead{
\colhead{Redshift} & \colhead{Aperture} & 
\colhead{NUV} & \colhead{W1} & 
\colhead{W2} & \colhead{W3} & 
\colhead{100 $\mu$m} & \colhead{160 $\mu$m} & 
\colhead{250 $\mu$m} & \colhead{350 $\mu$m} & \colhead{500 $\mu$m} \\ 
\colhead{} & \colhead{(Mpc)} & \colhead{($\mu$Jy)} & \colhead{($\mu$Jy)} & 
\colhead{($\mu$Jy)} & \colhead{($\mu$Jy)} & \colhead{(mJy)} & \colhead{(mJy)} & \colhead{(mJy)} & \colhead{(mJy)} & \colhead{(mJy)}
} 

\startdata
0.58 & r$\leq$0.25 & $2.42_{-0.29}^{+0.30}$ & $444\pm 10$ & $284.1_{-7.6}^{+7.5}$ & $364_{-26}^{+31}$ & $11.0\pm 4.7$ & $20.3\pm 3.8$ & $21.1_{-1.1}^{+1.3}$ & $11.79_{-0.81}^{+0.92}$ & $5.37_{-0.47}^{+0.53}$ \\
  & 0.25$\leq$r$\leq$0.50 & $3.48_{-0.63}^{+0.61}$ & $398\pm 14$ & $262\pm 12$ & $475_{-51}^{+50}$ & $28.0_{-7.4}^{+7.3}$ & $28.3_{-5.7}^{+5.6}$ & $27.4_{-1.9}^{+1.8}$ & $16.2_{-1.3}^{+1.2}$ & $7.19_{-0.73}^{+0.68}$ \\
  & 0.50$\leq$r$\leq$0.75 & $<1.8$ & $170\pm 16$ & $126\pm 15$ & $145\pm 66$ & $<17$ & $14.3\pm 6.9$ & $17.2_{-2.4}^{+2.3}$ & $9.1_{-1.6}^{+1.5}$ & $5.05_{-0.97}^{+0.95}$ \\
  & 0.75$\leq$r$\leq$1.00 & $<2.2$ & $59\pm 19$ & $52\pm 19$ & $<150$ & $<21$ & $<17$ & $9.3\pm 2.5$ & $5.5\pm 1.7$ & $3.3\pm 1.1$ \\
\hline
0.72 & r$\leq$0.25 & $2.41\pm 0.27$ & $358.7_{-8.3}^{+7.8}$ & $214.3_{-6.0}^{+5.8}$ & $407_{-31}^{+30}$ & $<7.9$ & $19.6\pm 3.1$ & $20.0_{-1.1}^{+1.0}$ & $12.95_{-0.76}^{+0.72}$ & $5.75_{-0.50}^{+0.48}$ \\
  & 0.25$\leq$r$\leq$0.50 & $2.64_{-0.56}^{+0.49}$ & $314_{-10}^{+11}$ & $196.5_{-9.0}^{+9.2}$ & $445_{-47}^{+40}$ & $<12$ & $12.1\pm 4.7$ & $24.6_{-1.5}^{+1.3}$ & $16.33_{-1.09}^{+0.97}$ & $7.19_{-0.63}^{+0.56}$ \\
  & 0.50$\leq$r$\leq$0.75 & $1.45\pm 0.72$ & $105\pm 11$ & $75\pm 11$ & $203\pm 52$ & $<14$ & $<12$ & $10.9\pm 1.7$ & $8.7\pm 1.2$ & $4.01_{-0.75}^{+0.74}$ \\
  & 0.75$\leq$r$\leq$1.00 & $<1.8$ & $45\pm 14$ & $37\pm 13$ & $<130$ & $<18$ & $<14$ & $5.7\pm 2.1$ & $4.3\pm 1.5$ & $2.34\pm 0.87$ \\
\hline
0.9 & r$\leq$0.25 & $1.77_{-0.21}^{+0.24}$ & $320.4_{-8.1}^{+7.6}$ & $218.3_{-6.6}^{+6.3}$ & $287_{-26}^{+27}$ & $<8.1$ & $14.0\pm 3.3$ & $19.4_{-1.3}^{+1.2}$ & $15.0_{-1.2}^{+1.0}$ & $6.84_{-0.65}^{+0.58}$ \\
  & 0.25$\leq$r$\leq$0.50 & $2.13_{-0.48}^{+0.41}$ & $251\pm 11$ & $170.7_{-9.4}^{+9.7}$ & $275_{-46}^{+39}$ & $<12$ & $13.4\pm 4.6$ & $25.0_{-1.8}^{+1.5}$ & $17.3_{-1.3}^{+1.0}$ & $8.55_{-0.70}^{+0.56}$ \\
  & 0.50$\leq$r$\leq$0.75 & $1.52_{-0.57}^{+0.56}$ & $110\pm 11$ & $93\pm 11$ & $138\pm 50$ & $<15$ & $<11$ & $15.5_{-1.8}^{+1.7}$ & $11.6_{-1.3}^{+1.2}$ & $5.54_{-0.74}^{+0.72}$ \\
  & 0.75$\leq$r$\leq$1.00 & $<1.4$ & $44\pm 13$ & $52\pm 14$ & $<110$ & $<18$ & $<14$ & $10.7\pm 2.0$ & $7.5\pm 1.4$ & $3.85\pm 0.81$ \\
\hline
1.1 & r$\leq$0.25 & $1.26_{-0.19}^{+0.21}$ & $258.9_{-5.8}^{+5.6}$ & $216.2_{-5.2}^{+5.0}$ & $308_{-26}^{+24}$ & $<7.1$ & $19.1\pm 3.1$ & $22.8_{-1.1}^{+1.0}$ & $18.11_{-0.87}^{+0.81}$ & $9.17_{-0.58}^{+0.55}$ \\
  & 0.25$\leq$r$\leq$0.50 & $1.50_{-0.44}^{+0.36}$ & $188.9_{-7.5}^{+7.8}$ & $156.6_{-7.3}^{+7.5}$ & $284_{-37}^{+32}$ & $<11$ & $18.7\pm 4.0$ & $21.7_{-1.5}^{+1.3}$ & $18.0_{-1.2}^{+1.0}$ & $8.84_{-0.69}^{+0.57}$ \\
  & 0.50$\leq$r$\leq$0.75 & $<1.0$ & $69.7\pm 8.7$ & $63.7\pm 8.4$ & $<83$ & $<13$ & $<9.6$ & $9.3\pm 1.4$ & $8.19_{-0.96}^{+0.95}$ & $4.56_{-0.64}^{+0.63}$ \\
  & 0.75$\leq$r$\leq$1.00 & $<1.2$ & $29\pm 11$ & $26\pm 11$ & $<98$ & $<16$ & $<11$ & $3.8\pm 1.6$ & $4.5\pm 1.2$ & $1.81\pm 0.70$ \\
\hline
1.3 & r$\leq$0.25 & $1.08_{-0.19}^{+0.20}$ & $198.4_{-4.9}^{+4.5}$ & $197.1_{-4.4}^{+4.2}$ & $267_{-31}^{+23}$ & $<6.9$ & $16.5\pm 2.7$ & $25.43_{-1.01}^{+0.89}$ & $22.06_{-0.93}^{+0.79}$ & $11.96_{-0.65}^{+0.56}$ \\
  & 0.25$\leq$r$\leq$0.50 & $1.34_{-0.48}^{+0.35}$ & $131.6_{-6.0}^{+7.0}$ & $143.9_{-5.9}^{+6.9}$ & $219_{-45}^{+32}$ & $<9.5$ & $8.2\pm 3.8$ & $21.6\pm 1.3$ & $18.9_{-1.2}^{+1.1}$ & $10.70_{-0.72}^{+0.70}$ \\
  & 0.50$\leq$r$\leq$0.75 & $<0.95$ & $40.0\pm 7.2$ & $52.9\pm 7.4$ & $128_{-40}^{+39}$ & $<11$ & $<9.4$ & $10.1\pm 1.3$ & $9.36_{-0.95}^{+0.96}$ & $5.46_{-0.60}^{+0.61}$ \\
  & 0.75$\leq$r$\leq$1.00 & $1.32\pm 0.58$ & $26.8\pm 9.0$ & $33.8\pm 9.4$ & $<96$ & $<14$ & $<11$ & $5.6\pm 1.6$ & $5.2_{-1.0}^{+1.1}$ & $2.88\pm 0.65$ \\
\hline
1.53 & r$\leq$0.25 & $0.85_{-0.23}^{+0.22}$ & $140.3\pm 4.1$ & $168.9_{-4.6}^{+4.7}$ & $190_{-26}^{+25}$ & $<8.2$ & $19.5\pm 3.4$ & $26.4\pm 1.1$ & $23.6\pm 1.1$ & $13.93_{-0.75}^{+0.80}$ \\
  & 0.25$\leq$r$\leq$0.50 & $<0.71$ & $96.5_{-6.5}^{+5.6}$ & $121.3_{-7.8}^{+6.5}$ & $168_{-54}^{+35}$ & $<11$ & $10.6\pm 4.4$ & $21.0_{-1.9}^{+1.5}$ & $18.4_{-1.7}^{+1.3}$ & $11.74_{-1.16}^{+0.83}$ \\
  & 0.50$\leq$r$\leq$0.75 & $<1.0$ & $38.3\pm 7.5$ & $51.4\pm 8.5$ & $<93$ & $<14$ & $<10$ & $10.6\pm 1.5$ & $10.1_{-1.1}^{+1.0}$ & $6.25_{-0.70}^{+0.66}$ \\
  & 0.75$\leq$r$\leq$1.00 & $<1.3$ & $<18$ & $36\pm 11$ & $<110$ & $<17$ & $<12$ & $7.5\pm 1.7$ & $5.1\pm 1.2$ & $3.59_{-0.71}^{+0.70}$ \\
\hline
1.77 & r$\leq$0.25 & $0.77_{-0.28}^{+0.27}$ & $109.1_{-5.0}^{+4.7}$ & $139.0_{-5.7}^{+5.4}$ & $149_{-31}^{+28}$ & $<9.7$ & $9.9\pm 4.0$ & $23.7_{-1.5}^{+1.3}$ & $22.1_{-1.2}^{+1.1}$ & $13.63_{-0.94}^{+0.96}$ \\
  & 0.25$\leq$r$\leq$0.50 & $<0.98$ & $70.2_{-7.4}^{+6.9}$ & $89.2_{-9.1}^{+8.5}$ & $<62$ & $<12$ & $<9.9$ & $15.3_{-2.2}^{+1.8}$ & $15.4_{-2.0}^{+1.6}$ & $10.4_{-1.5}^{+1.1}$ \\
  & 0.50$\leq$r$\leq$0.75 & $<1.4$ & $29.3\pm 8.5$ & $26.4\pm 9.9$ & $<100$ & $<16$ & $<11$ & $6.4\pm 1.6$ & $7.7\pm 1.2$ & $5.87_{-0.83}^{+0.78}$ \\
  & 0.75$\leq$r$\leq$1.00 & $<1.9$ & $<21$ & $<25$ & $<120$ & $<22$ & $<16$ & $<3.9$ & $3.3\pm 1.4$ & $2.81\pm 0.81$ \\
\hline
2.07 & r$\leq$0.25 & $<0.64$ & $93.3_{-6.1}^{+5.3}$ & $110.8_{-7.3}^{+6.3}$ & $235_{-43}^{+37}$ & $<12$ & $16.3\pm 4.4$ & $21.3_{-1.5}^{+1.4}$ & $19.3_{-1.4}^{+1.3}$ & $12.3\pm 1.1$ \\
  & 0.25$\leq$r$\leq$0.50 & $1.33_{-0.66}^{+0.56}$ & $65.3_{-8.4}^{+8.6}$ & $82\pm 11$ & $117_{-51}^{+50}$ & $<17$ & $<12$ & $18.0_{-2.5}^{+1.9}$ & $16.1_{-2.3}^{+1.6}$ & $10.7_{-1.6}^{+1.1}$ \\
  & 0.50$\leq$r$\leq$0.75 & $<1.6$ & $31\pm 10$ & $35\pm 13$ & $<130$ & $<20$ & $<16$ & $9.8\pm 2.4$ & $8.5\pm 1.6$ & $5.47_{-1.03}^{+0.96}$ \\
  & 0.75$\leq$r$\leq$1.00 & $<2.0$ & $<27$ & $<32$ & $<150$ & $<25$ & $<19$ & $5.6\pm 2.7$ & $4.6\pm 1.7$ & $2.5\pm 1.1$ \\
\hline
2.37 & r$\leq$0.25 & $0.66\pm 0.28$ & $77.2_{-5.0}^{+4.5}$ & $94.5_{-6.5}^{+5.6}$ & $207_{-40}^{+35}$ & $<11$ & $25.0_{-4.8}^{+4.7}$ & $19.2_{-1.6}^{+1.3}$ & $17.0_{-1.4}^{+1.1}$ & $10.51_{-1.09}^{+0.96}$ \\
  & 0.25$\leq$r$\leq$0.50 & $<0.85$ & $50.1_{-7.6}^{+7.7}$ & $59.4_{-8.9}^{+9.0}$ & $<97$ & $<16$ & $20.2_{-6.4}^{+6.1}$ & $14.6_{-2.4}^{+2.0}$ & $13.3_{-2.2}^{+1.6}$ & $8.7_{-1.6}^{+1.1}$ \\
  & 0.50$\leq$r$\leq$0.75 & $<1.4$ & $<21$ & $<23$ & $<120$ & $<19$ & $<15$ & $9.9\pm 2.3$ & $8.9_{-1.6}^{+1.5}$ & $5.24_{-1.00}^{+0.94}$ \\
  & 0.75$\leq$r$\leq$1.00 & $<1.8$ & $<26$ & $<30$ & $<150$ & $<24$ & $<19$ & $<5.0$ & $4.4\pm 1.8$ & $2.5\pm 1.1$ \\
\enddata
\end{deluxetable*}

\bibliography{ref}{}
\bibliographystyle{yahapj}



\end{document}